\documentclass[a4paper,11pt]{article}
\usepackage{aaskaiid}
\usepackage{orcidlink}
\def\ltsima{$\; \buildrel < \over \sim \;$}
\def\simlt{\lower.5ex\hbox{\ltsima}}
\def\gtsima{$\; \buildrel > \over \sim \;$}
\def\simgt{\lower.5ex\hbox{\gtsima}}

\title{Exploring Tidal Disruption Events with SKA and VLBI: Unveiling the Mystery of Black Hole Feeding and Outflows}
\ShortTitle{TDEs with the SKA and VLBI}

\author[1,\dagger]{Xinwen Shu\orcidlink{0000-0002-7020-4290}}
\ShortName{Shu et al.} % shortened name list for header 
\author[2]{Guobin Mou\orcidlink{0000-0002-0092-7944}}
\author[3]{Tao An\orcidlink{0000-0003-4341-0029}}
\author[4]{S. Komossa\orcidlink{0000-0003-4183-4215}}
\author[5,\dagger]{Miguel P\'erez-Torres\orcidlink{0000-0001-5654-0266}}
\author[6]{Weihua Lei\orcidlink{0000-0003-3440-1526}}
\author[1]{Luming Sun\orcidlink{0000-0002-7223-5840}}
\author[3]{Ning Jiang\orcidlink{0000-0002-7152-3621}}
\author[3]{Tinggui Wang\orcidlink{0000-0002-1517-6792}}
\author[7]{Chichuan Jin\orcidlink{0000-0002-2006-1615}}
\author[8]{Jun Yang\orcidlink{0000-0002-2322-5232}}

\affiliation[1]{Department of Physics, Anhui Normal University, Wuhu, Anhui 241002, China}
\emailAdd{xwshu@ahnu.edu.cn}
\affiliation[2]{Department of Physics and Institute of Theoretical Physics, Nanjing Normal University, Nanjing 210023, China}
\emailAdd{gbmou@njnu.edu.cn}
\affiliation[3]{Department of Astronomy, University of Science and Technology of China, Hefei, Anhui 230026, China}
\emailAdd{antao@shao.ac.cn}
\affiliation[4]{Max-Planck-Institut f$\ddot{u}$r Radioastronomie, Auf dem H$\ddot{u}$gel 69, 53121 Bonn, Germany}
\emailAdd{skomossa@mpifr.de}
\affiliation[5]{Instituto de Astrof\'isica de Andaluc\'ia, Consejo Superior de Investigaciones Cient\'ificas (CSIC), Glorieta de la Astronom\'ia s/n, E-18008, Granada, Spain}
\emailAdd{torres@iaa.csic.es}
\affiliation[6]{Department of Astronomy, Huazhong University of Science and Technology, Wuhan  430074, China}
\emailAdd{leiwh@hust.edu.cn}
\affiliation[7]{National Astronomical Observatories, Chinese Academy of Sciences, Beijing 100101, China}
\affiliation[8]{Onsala Space Observatory, SE-439 92 Onsala, Sweden}

\affiliation[\dagger]{Chapter co-ordinator}

\abstract{
{\bf Abstract:} Tidal disruption events (TDEs) probe the birth and evolution of black‑hole accretion flows and jets on human timescales. Radio emission traces shocks and outflows from non‑jetted (``thermal'') TDEs and powerful relativistic jets in the rare ``jetted'' class. SKA‑Mid, phased for VLBI and used together with global networks, will deliver sub‑mas imaging, tens‑of‑$\mu$as astrometry, and $\mu$Jy sensitivity, enabling: (i) proper‑motion measurements that discriminate off‑axis relativistic jets from sub‑relativistic winds; (ii) resolved morphologies and magnetic‑field diagnostics via polarimetry; and (iii) precise nuclear localization to distinguish SMBH vs. IMBH and to reveal recoil or binary systems. SKA's wide frequency coverage (0.35--15.4~GHz) and 1‑h continuum sensitivities of $\approx 3-10 \mu$Jy~beam$^{-1}$, together with multi‑beam tied‑array VLBI and a transient buffer for rapid triggers, are transformational. Rubin/LSST, Einstein Probe, and SVOM will increase TDE alerts to tens--hundreds per year, and late‑time radio flares appear common ($\simgt40\%$), ensuring rich SKA--VLBI samples. We provide observing strategies, detection forecasts, and predictions (e.g., $\geq 5$ proper‑motion detections of jetted/off‑axis TDEs per year; routine $\mu$as‑level core‑shift constraints). This program will establish TDEs as laboratories for exploring jet launching, particle acceleration (including neutrinos), black hole accretion history and demographics, and properties of circumnuclear medium. 

}

\begin{document}
\maketitle

\section{Introduction}
Tidal disruption events (TDEs) represent a captivating intersection of astrophysics, occurring when a star is drawn too close to a supermassive black hole (SMBH) and is torn apart by tidal forces \citep{Rees1988},  producing bright, multi‑wavelength flares \citep{Gezari2021, Komossa2015}. These events provide a unique opportunity to study black hole feeding mechanisms, accretion processes, outflows, relativistic jets, and the environments surrounding {SMBHs \citep[][in this Book]{TaoAn03.2026.SKA}. }
%(An et al. in this Book). 
The Square Kilometre Array (SKA), in conjunction with Very Long Baseline Interferometry (VLBI), offers unprecedented capabilities to investigate the radio emissions associated with TDEs, addressing key questions in the field:

(1) Jet formation and evolution: Incipient jet kinematics (proper motion via VLBI) and spectral energy distribution (SED) evolution during the emergent phase. 
{The prospects of using SKA+VLBI to address this question will be presented in Sections 2, 3, 5 and 7.}

(2) Delayed Emission Physics: Temporal correlation between accretion-state transitions (e.g., disk-jet coupling timescales) traced by optical and X-ray observations, and delayed radio flares via high-cadence multi-epoch monitoring. 
{The unique role of SKA+VLBI observations in uncovering the origins of delayed radio flares will be presented in Section 3.}

(3) Properties of circumnuclear medium (CNM): Statistical constraints on the CNM inhomogeneities at scales of (sub-)parsecond (pc), through population-level outflow-CNM interaction studies (in comparison with {dedicated hydrodynamic simulations). 
SKA+VLBI observations will enable to diagnose the properties of smooth CNM and dense clouds (Section 4).}

(4) Demographic Completeness: Unbiased detection of rare TDE subtypes (high-z events, off-axis jets, 
%IMBH- and SMBHB-driven disruptions) 
and systems involving intermediate-mass black hole (IMBH) and SMBH binary (SMBHB)) 
across diverse host galaxy environments. 
{The scientific applications of SKA+VLBI for unbiased detections of TDEs will be presented in Section 5}. 

(5) Multi-messenger Associations: Understanding the driving mechanisms that produce the high energy neutrinos in TDEs.  
{SKA+VLBI observations are able to localize precisely the radio emitting region and place a rigorous constraint on the energetics relevant to the production of neutrinos (Section 6).}

%\citep{braun2019anticipatedperformancesquarekilometre}.

\section{Relativistic jet formation and evolution}

\subsection{Background}

The condition for relativistic jet formation and its connection with 
accretion onto SMBHs is largely unclear.
Stellar TDE by SMBH lead to dramatic changes in accretion rate by orders of magnitude \citep[e.g.,][]{Rees1988, Komossa2015}, presenting a unique laboratory to study jet formation.
Despite more than a hundred TDEs discovered so far, only four have been observed to launch relativistic jets; three are discovered in the hard X-rays, {including Swift J1644+5734 \citep{Burrows2011} , Swift J2058.4+0516 \citep{Cenko2012}, and Swift J1112-8238 \citep{Brown2015}},  
%including the prototype Swift J1644+5734 \citep{Burrows2011}  
and one in the optical, {AT2022cmc} \citep{Andreoni2022}.
Jetted TDEs could be misidentified as long Gamma-ray bursts (GRBs) or X-ray flashes \citep{Tchekhovskoy2014, Shu2025, Li2026}, making the true abundance of jetted TDEs largely uncertain.
In addition, it remains puzzling why only a small fraction ($<$1\%) of TDEs launch successful jets.
%(e.g., Sun et al. 2015; Yao et al. 2024)
Possible explanations include that jet formation requires special conditions on the SMBH spin and magnetic field 
%(e.g., Giannios \& Metzger 2011; 
\citep[e.g.,][]{Tchekhovskoy2014}, 
the jet is chocked by the disk wind in most TDEs \citep{Teboul2023, Lu2024}, or the large off-axis viewing angle of jet 
\citep{Matsumoto2023}.
However, the limited data on jetted TDEs prevents adequate testing of these explanations.
%{\bf Thus, there is a pressing need to enlarge the sample of jetted TDEs.}

%\begin{figure}[h]
%    \centering
%	\includegraphics[width=0.3\columnwidth]{SKAO Pictorial mark-01.png}
%    \caption{Example figure. Place caption below table.}
%    \label{fig:example_figure}
%\end{figure}

%Referring to Fig.~\ref{fig:example_figure}.

\subsection{SKA-VLBI capability, observing strategy, and expected results}

%\begin{equation}
%    x=\frac{-b\pm\sqrt{b^2-4ac}}{2a}.
%	\label{eq:quadratic}
%\end{equation}

%Referring to equation~(\ref{eq:quadratic}).

\begin{figure}
\centering
\includegraphics[width=0.95\columnwidth]{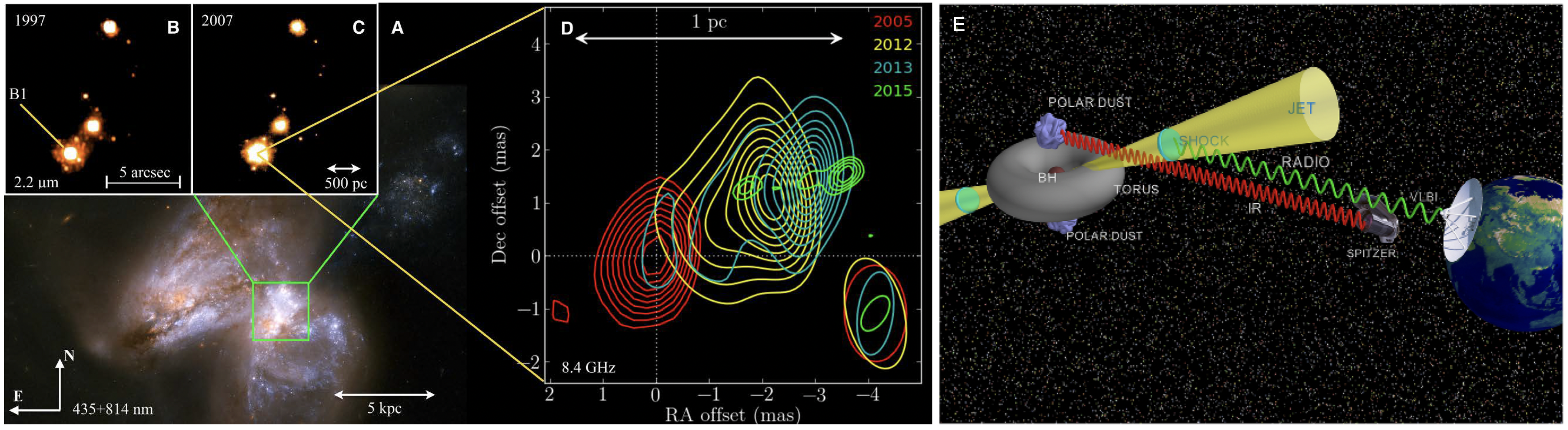}
 \caption{ The left panels show the near-infrared outburst of the TDE Arp 299-B AT1 ({\bf B} and {\bf C}) and the color-composite optical image of its host galaxy ({\bf A}). 
 The evolution of the radio morphology as imaged with VLBI at 8.4 GHz \citep{Mattila2018} is shown in panel {\bf D}. 
 It can be seen that an initially unresolved radio source develops into a resolved jet structure a few years after the explosion, which is viewed 
 with a slightly off-axis angle ($\theta_{\rm obs}=25^{\circ}-35^{\circ}$, {\bf E}).
 %the snapshots of gas density and energy density of relativistic electrons during the outflow-CNM/cloud interaction (in Gaussian-cgs units), as well as the synthetic radio spectra ($d_L = 100$ Mpc). The spectral panel displays both the BS+FS component and the FS component alone, from which the impact and the rapid variability feature of the BS can be observed. The shaded regions represent the frequency range covered by SKA-Mid. The right panels show the synthetic radio maps, with axes labeled in milliarcseconds. The three epochs correspond to the phases before the BS emerges (3 yr, when only the forward shock is present), the early stage (4 yr) and the late stage (6 yr) of the outflow-cloud interaction (\citealt{Mattila2018}). 
 } 
 \label{fig1jet}
\end{figure}

All four on-axis jetted TDEs exhibit luminous ($\nu L_\nu>10^{40}$ erg s$^{-1}$) and long-lived radio emission. 
However, no apparent superluminal motion has been detected in the known jetted TDEs with VLBI observations at a resolution of $\sim$1 mas, such as Swift J1644+5734 \citep{Yang2016}, possibly due to a very small viewing angle, a strong jet deceleration, or a too small physical scale. 
With the VLBI observations over a decade, \citet{Mattila2018} reported the discovery of a resolved and expanding radio jet ($\beta\approx0.3$c) in the dust-enshrouded TDE Arp 299-B AT1 at $z=0.01$, likely originating from a jet viewing slightly off-axis ($\theta_{\rm obs}=25^{\circ}-35^{\circ}$, Figure \ref{fig1jet}).

%[SK] 
The two high-energy instruments aboard the Space Variable Objects Monitor \citep[SVOM;][]{Wei2016} space mission, the ECLAIRs and GBM (4-150 keV and 15-5000 keV, respectively) are characterized by a large field of view and are optimally suited for the detection of new jetted TDEs.  The space mission Einstein Probe \citep[EP;][]{Yuan2025} will additionally detect soft X-ray TDEs that may launch relativistic radio-emitting outflows. Both missions stand out due to their rapid alert time after the discovery of new transients within minutes, ensuring quick follow-ups in the radio regime and therefore probing the very early phases of jet launching. 
Radio follow-ups of extraordinary bright TDEs %($L>10^{45}$ erg/s) 
discovered in time-domain X-ray surveys from SVOM and EP, optical surveys from LSST, or directly by radio surveys with SKA have the potential to resolve jet emission and probe the dynamics of jet evolution, especially in the early ultra-relativistic phase.

Over its past one-year operation, EP has detected a few jetted TDE candidates in real time \citep{Shu2025, Li2026}. 
%{\bf Given the current progress achieved by ZTF and EP, 
{We anticipate that at least 1-2 jetted TDEs per year would be detected in the coming LSST, EP and SVOM surveys, which are ideal targets for follow-up observations by SKA+VLBI.  
SKA-Mid Bands 2 (0.95–1.76 GHz) and 5 (4.6–15.4 GHz) provide the critical  SKA+VLBI frequency coverage for size measurements and astrometry studies. 
The intercontinental baselines imply a spatial resolution of $\theta_{\rm beam}\simeq\lambda/B_{\rm max}$ $\approx$1 mas at 8 GHz and $\approx$0.5 mas at 15 GHz. 
The naturally weighted map $rms$ for a global multi-baseline imaging including SKA-Mid reaches a few $\mu$Jy/beam, e.g., $\sim$3$\mu$Jy/beam in $\sim$1 hour integration, depending on the actual antennas in operation and processed bandwidth. 
The positional uncertainties of $\sigma_{\rm pos}\approx\theta_{\rm beam}/(\rm 2\times SNR)$
, where SNR is the signal-to-noise ratio, suggest that tens of $\mu$as relative astrometry can be achieved for mJy-level sources.  
%For jetted TDEs with a flux of $\sim$1 mJy at $\sim$10 GHz, 
Once triggered as jetted TDE candidates from optical and X-ray surveys, we envisage that the SKA observations in commensal mode during the first a few days of evolution will determine whether the radio flare is detectable or not. If the target is bright enough in the SKA-Mid bands (i.e., $\simgt$0.5 mJy), 3-4 long ($\simgt$8 h) SKA+VLBI tracks can be scheduled, spanning a few weeks to months post trigger. 
%, enabling to constrain the proper motion and mas-scale morphology of a newly launched jet. 
%SKA+VLBI will allow for resolving any structures at a (sub-)pc scale 
For jetted TDEs at z$\simlt$0.1, 
%for a resolution range $\sim$0.5-1 mas, which is critical to 
SKA+VLBI will enable to directly resolve the radio emission components at pc scales, or measure the proper motion,  %, providing a smoking-gun 
providing a compelling evidence for the presence of a radio jet. 
%the observed radio jet launched by a TDE. 
In addition, multi-frequency SKA+VLBI observations, especially at Bands 5a and 5b, will provide the precise position (tens of $\mu$as astrometry) of the compact core as a function of observing frequency. 
The measured offset between frequency pairs yields a core-shift parameter, 
which can be used to estimate the magnetic field strength of the jet-emitting region \citep[e.g.,][]{Sokolovsky2011}. 
%under modest assumptions about geometry and Doppler factor \citep[e.g.,][]{Sokolovsky2011}. 
The core-shift-derived magnetic field strength will be important to understand the physical 
conditions for jet launching. 
%validate the simulation results of FS and BS evolution. 
}

%[SK] We may highlight here the previous spatial resolution of the powerful jet of SwiftJ1644 \citep{Yang2016}, and discuss future application to newly discovered jets (e.g., measurement of superluminal motion), but perhaps also the application to follow the late-time evolution of some of the existing jetted TDEs.  SwiftJ1644 itself unfortunately is at rather high z. )

%\subsection{Tables}

%\begin{table}[h]
%	\centering
%	\caption{Example table. Place caption above each table.
%	Remember to define the quantities, symbols and units used.}
%	\label{tab:example_table}
%	\begin{tabular}{lccr} % four columns, alignment for each
%		\hline
%		A & B & C & D\\
%		\hline
%		1 & 2 & 3 & 4\\
%		2 & 4 & 6 & 8\\
%		3 & 5 & 7 & 9\\
%		\hline
%	\end{tabular}
%\end{table}

%Referring to Table~\ref{tab:example_table}.

\section{Delayed emission physics and non-relativistic outflows }
%\subsection{Background }
 %Reviewing the results of literature discovery of delayed radio flares in TDEs, some of which show double or more brightenings. The origins of delayed radio emission can be reviewed as well, i.e., pros and cons between different models. 

In comparison with jetted TDEs, radio emission from thermal TDEs has lower luminosities, {possibly dominated by non-relativistic outflows}. %, and appears to be ubiquitous. 
{Thermal TDEs refer to those flares in optical/UV and X-rays that can be described by a blackbody emission, which is distinct from the non-thermal emission from jetted TDEs. 
The low-luminosity radio emission in thermal TDEs appears to be ubiquitous \citep{Alexander-et-al.(2016)} 
}. 
Recently, \cite{Cendes2024} present the results from radio follow-up observations of 23 optical TDEs and find that $\sim 40\%$ of them exhibit radio brightening hundreds to thousands of days after the discovery. %, the origin remains unclear.   %\cite{Cendes..2024} propose delayed 
%More interestingly, a few TDEs display %Monitoring the flux evolution of the delayed radio emission reveals 
%steeply rising light curves in the delayed radio emission \citep{Horesh..2021a, Cendes..2022, Zhang..2024, Sfaradi2025},
%These flares exhibit a flux density jump from nondetection to detection, 
%requiring a temporal power-law index steeper than $t^4$—a behavior inconsistent with the predictions of standard model that a single outflow (either relativistic or sub-relativistic) is launched into the smooth CNM promptly following the TDE. 
%The origins of delayed, fast-rising radio emission are still largerly unknown. 
%outflow-CNM models.
The nature of the late-time radio brightening in TDEs is still under debate, 
and several scenarios have been proposed, % to explain the behaviours of the late-time radio brightening, 
including decelerating and expansion of an off-axis jet launched at the time of disruption \citep[e.g.,][]{Matsumoto2023}, 
a delayed launching of the outflow perhaps from delayed disk formation \citep{Horesh2021, Cendes2024}, 
%propagation of the outflow in an inhomogeneous medium \citep[e.g.,][]{Matsumoto2024}, 
changes in the structure and density profile of CNM \citep[e.g.,][]{Mou2022, Matsumoto2024},
%\textcolor{black}{outflow--torus/cloud interaction \citep{Mou2021, Mou2022, Zhuang2024}},
and break out of the choked precessing jets from the wind envelop \citep{Teboul2023, Lu2024}. 

% \subsection{Synergy with LSST, EP and SVOM}

\begin{figure}
\centering
\includegraphics[width=0.95\columnwidth]{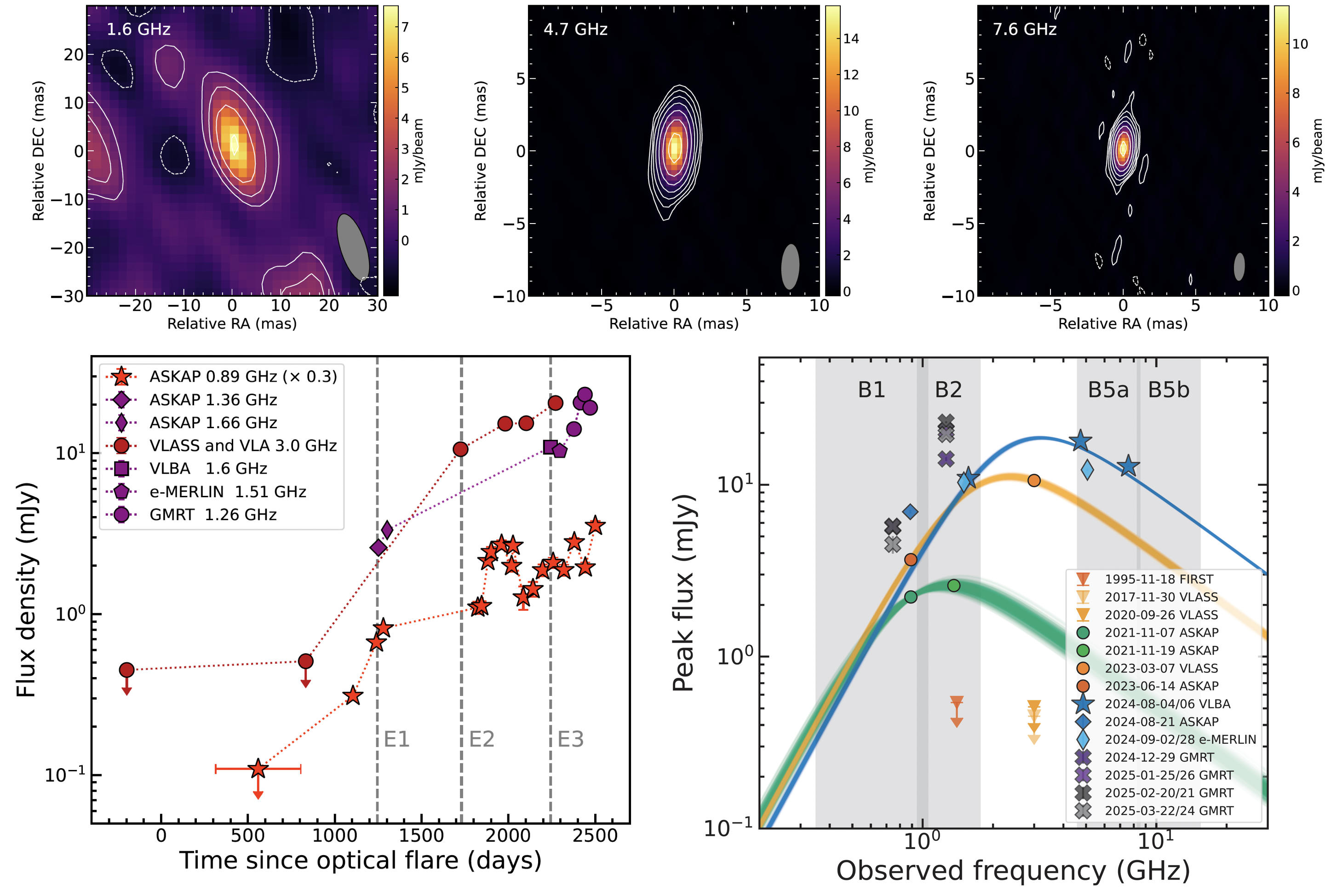}
 \caption{ Natural-weighted images of VLBA observations of AT2018cqh at three frequencies (upper panel), 
 which reveal a compact radio emission, unresolved at a scale of $\simlt$0.13 pc \citep{yang2025}. 
 Lower panel shows the radio light curves at 0.89 GHz, 1.3-1.6 GHz and 3 GHz (left), and the radio SEDs 
 over three epochs (right).  %that have quasi-simultaneous observations at different frequencies (right).   
The green, orange, and blue lines represent the best fits to each SED, 
{which are the model realizations of MCMC fittings. 
%from MCMC modeling. 
The shaded regions represent the frequency range covered by SKA-Mid. 
There is a steady rise in both the peak flux
density and frequency between the three epochs, which have rarely been observed in TDEs.} 
 %It can be seen that an initially unresolved radio source develops into a resolved jet structure a few years after the explosion (D), which is viewed 
 %with a slightly off-axis angle ($\theta_{\rm obs}=25^{\circ}-35^{\circ}$, E).
 %the snapshots of gas density and energy density of relativistic electrons during the outflow-CNM/cloud interaction (in Gaussian-cgs units), as well as the synthetic radio spectra ($d_L = 100$ Mpc). The spectral panel displays both the BS+FS component and the FS component alone, from which the impact and the rapid variability feature of the BS can be observed. The shaded regions represent the frequency range covered by SKA-Mid. The right panels show the synthetic radio maps, with axes labeled in milliarcseconds. The three epochs correspond to the phases before the BS emerges (3 yr, when only the forward shock is present), the early stage (4 yr) and the late stage (6 yr) of the outflow-cloud interaction (\citealt{Mattila2018}). 
 } 
 \label{fig2outflow}
\end{figure}

VLBI observations will be able to test directly the scenario of an off-axis jet for the origin of delayed radio flares. 
In the off-axis scenario, assuming that the relativistic jet travels roughly perpendicularly to our line of sight at the speed of light for a typical TDE at a distance of $\sim$100 Mpc, the proper motion of the radio source is $\sim$0.6 mas/year. 
We propose to use SKA+VLBI {Band 5} to observe a sample of optical and/or X-ray thermal TDEs with late-time radio brightening, especially those with steeply rising light curves \citep[e.g.][]{Cendes2022}. 
%Given that the radio signal is rising 
%We expect that SKA+VLBI observations 
{3-4 long ($\simgt$8 h) SKA+VLBI observations over $\sim$1-3 years will enable to measure the proper motion since the rise in radio emission}, hence place an useful constraint on the off-axis scenario. 
%{\bf 3-4 long ($\simgt$8 h) SKA+VLBI tracks will be , 
If this scenario is confirmed, it implies that the jet has a very large energy of $\simgt10^{52}$ erg \citep{Beniamini2023, Matsumoto2023}, which will have interesting implications for the total energy released and provide new clues to resolve the ``missing energy" puzzle in TDEs \citep{Lu2018}. 
%the central engines in TDEs. 
Alternatively, a null result will rule it out, and favor the scenario of late launch of an outflow (Figure \ref{fig2outflow}). 
%In this case, 
On the other hand, VLBI observations are also crucial to constrain the CNM properties, for which we will discuss in more detail in next Section, by comparing with hydrodynamic simulations of the interaction of outflow with CNM {and/or dense clouds}. 

%Radio followup of normal optical TDEs discovered by LSST and X-ray TDEs by EP and/or SVOM. Temporal correlation between accretion-state transitions (e.g., disk-jet coupling timescales) and delayed radio flares via high-cadence multi-epoch monitoring. 
%The structure and morphology properties from SKA-VLBI observations can be used to constrain the origin of radio flares from outflow or low-luminosity jet. 
%SKA-VLBI observations of delayed and fasting rising radio emission in optical TDEs can test the scenario of off-axis jets, i.e., late-time emergence of jet-like components. 

% [SK] We may add SVOM: SVOM might preferentially detect the high-energy/gamma-ray emission of jetted TDEs that are interesting for radio follow-ups.

\section{Properties of the circumnuclear medium}

\subsection{Background }

%[GM] 
The origin of radio emissions in TDEs remains an open question in high-energy astrophysics. 
In most cases, radio emissions are thought to be generated by the forward/external shocks (FS) driven by outflow in the ambient CNM, while rare cases are thought to originate from internal shocks within the jet \citep{alexander2020SSRv,Pasham2018}. In such a forward shock scenario, using the multi-frequency data, one can fit the radio spectrum and determine the peak frequency $\nu_p$ and peak flux $F_{\nu_p}$. Following the energy equipartition method \citep{duran2013}, the size of the radiative zone and the total nonthermal energy and be calculated, which can be further used to estimate the radius and total amount of post-shock gas. Accordingly, the CNM density at different radii can be derived. 
%[LWH] 
Therefore, the radio emissions at different epochs can be used to construct the CNM density profile at pc or even sub-pc scales \citep{Alexander-et-al.(2016),Zhou2024}.

\subsection{Smooth CNM and dense clouds}

Recent radio observations have uncovered the diversity of radio afterglows in TDEs. 
An increasing number of radio afterglows exhibit a sharply rising flux and some show a derived shock radius that %decreases over time 
changes little with time 
when applying the FS model (e.g., \citealt{yang2025}). 
These features challenge the conventional FS model but are consistent with the bow shock (BS) model, in which the radio emission is produced by BS formed as the outflow impacts high-density gas. 
These findings suggest that BS and dense gas could exist in some TDEs, and more importantly, confirm that radio observations can be used to investigate the circumnuclear environment in galactic nuclei and to explore simultaneously the diffuse CNM and dense gas. 
%$radio observations can be used to investigate the circumnuclear environment in galactic nuclei and to explore the distribution of diffuse CNM and dense gas. 

%[LWH] 
On the other hand, the CNM density profile\,($n\propto R^{-k}$) provides key diagnostics of the accretion history of the SMBH. Currently, there are three special types of CNM density profiles: (1) Sgr $\rm A^{*}$-like\,($k=1$), namely, the CNM density profile consistent with Sgr $\rm A^{*}$ whose $k=1$ \citep{Baganoff03}, and AT2019dsg is one of the TDEs satisfying this distribution \citep{Stein2021}; (2) Bondi-like\,($k=3/2$), consistent with Bondi accretion whose $k=3/2$, and Sw J1644+57 is one of the TDEs satisfying this distribution \citep{Eftekhari18}; (3)\,ASASSN-14li-like ($k=5/2$), there are several TDEs satisfying this distribution, such as ASASSN-14li \citep{Alexander-et-al.(2016)}, CNSS J0019+00 \citep{Anderson20} and Arp 299-B AT1 \citep{Mattila2018}. Such a steep profile is not expected for spherically symmetric accretion, but appears in some models of super-Eddington accretion flows \citep[ZEBRAs]{Coughlin2014}. 
%The accretion history of a dormant SMBH especially at $z>1$ is of great interest.

%[ Basics of radio synchrotron emission produced in outflow-CNM interaction. 
%The importance of CNM distribution and structure, e.g., can be used to trace the accretion history of SMBH.  ]

%\subsection{New observations}
%Radio flux and SED evolution to constrain the density profile of the CNM, and distribution of dense clouds (if present). 

\subsection{Hydrodynamic simulations}

\begin{figure}
\centering
\includegraphics[width=0.35\columnwidth]{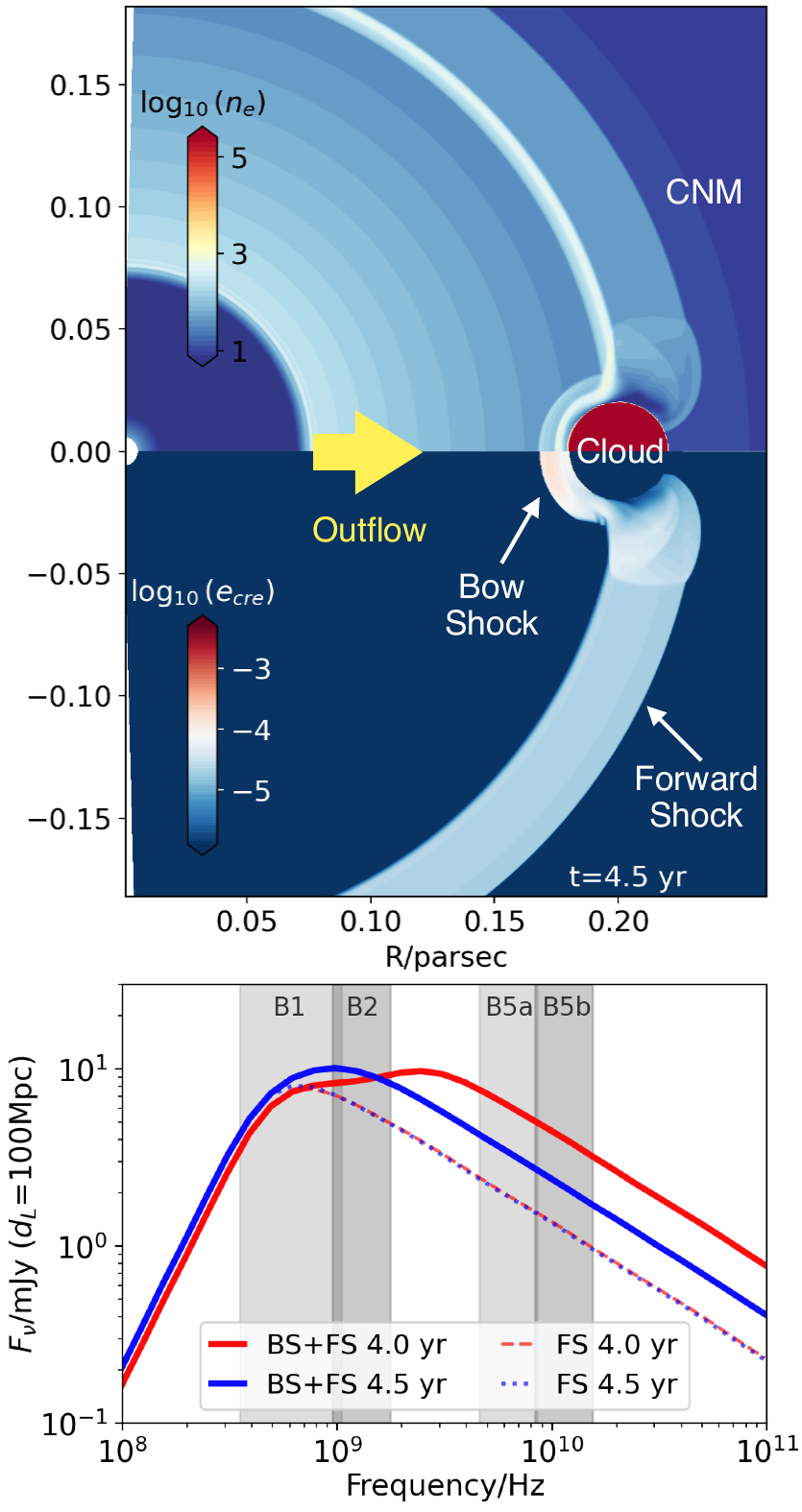}
\includegraphics[width=0.55\columnwidth]{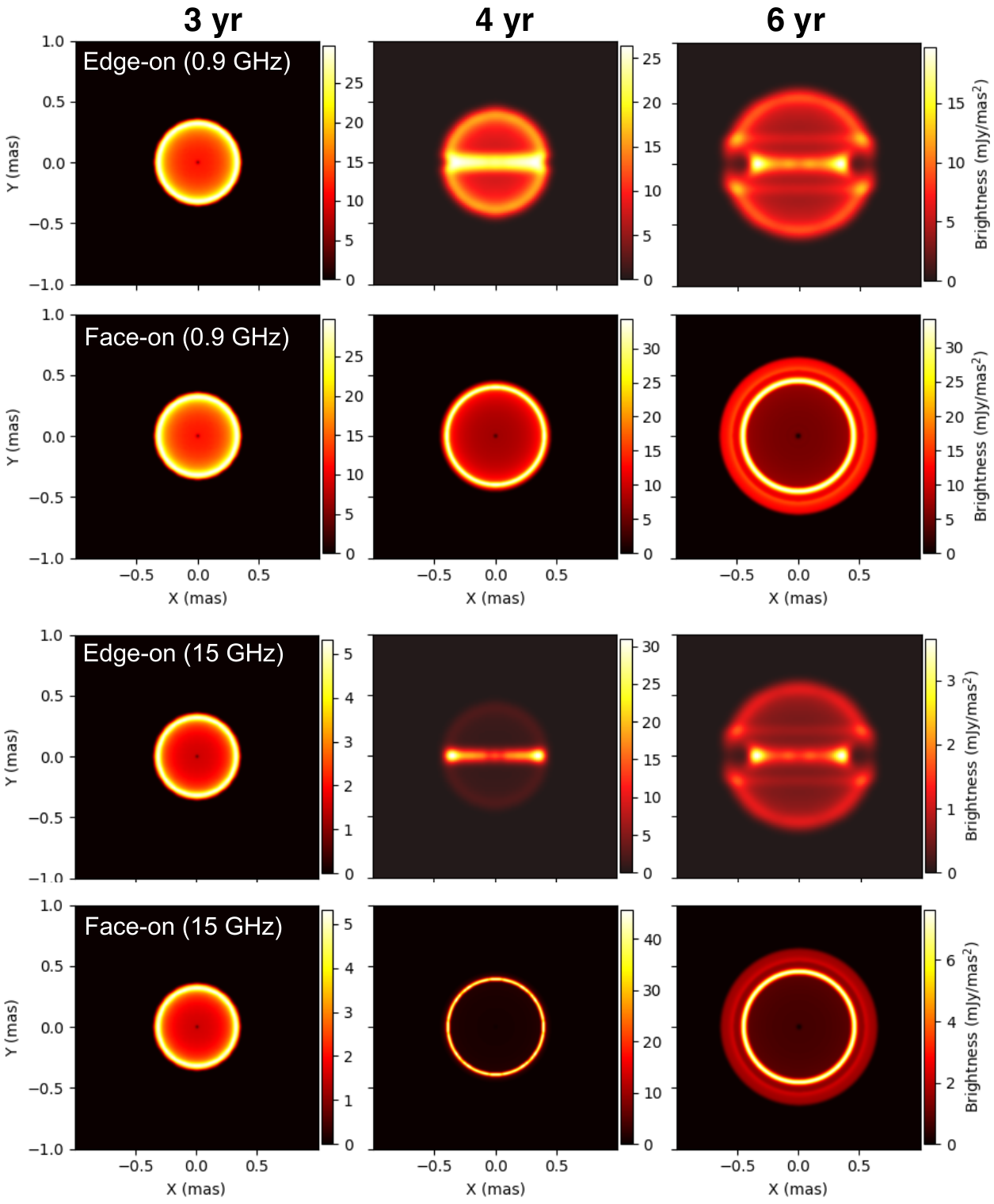}
 \caption{ The left panels show the snapshots of gas density and energy density of relativistic electrons during the outflow-CNM/cloud interaction (in Gaussian-cgs units), as well as the synthetic radio spectra ($d_L = 100$ Mpc). The spectral panel displays both the BS+FS component and the FS component alone, from which the impact and the rapid variability feature of the BS can be observed. The shaded regions represent the frequency range covered by SKA-Mid. The right panels show the synthetic radio maps, with axes labeled in milliarcseconds. The three epochs correspond to the phases before the BS emerges (3 yr, when only the forward shock is present), the early stage (4 yr) and the late stage (6 yr) of the outflow-cloud interaction \citep[][under review]{moushu2025}. } 
 \label{fig2map}
\end{figure}

%[Simulations of jet/outflow interaction with CNM or clouds \cite{Mou2022, yang2025}. ]
%[GM] 
Hydrodynamic simulations are crucial to understand the delayed physical process in the interaction of outflow with CNM and/or clouds. 
The high-energy electrons and magnetic fields in the FS are dispersed throughout the post-shock CNM, resulting in a relatively large spatial distribution region. Compared to the FS, the high-energy electrons and magnetic fields in the BS are concentrated at the shock front (the windward side of the cloud). These discrepancies lead to different spectra and temporal evolution of the radio emission from the BS compared to the FS. 
Recent two-fluid simulations incorporating the relativistic electrons as the second fluid have revealed the light curve patterns and spectral characteristics for both FS and BS, and prove the significant differences between these two types of {shocks \citep[][under review]{mou2025, moushu2025}.} For light curves at high frequencies ($\nu > \nu_p$), the flux variation of the FS is relatively gradual. 
%---the rising rate is generally slower than $t^1$, while the declining rate lies between $t^{-1}$ and $t^{-2}$, depending on the density profile index and the electron spectral index \cite{mou2025}. 
In contrast, the flux of the BS rises sharply with the onset of cloud impact and then declines rapidly as the interaction ends. 
The radio flux from the BS could exhibit short-term (monthly) variations when the outflow exhibits short-term fluctuations, while for the FS, its radio emission is hard to produce noticeable fluctuations on a timescale of months. 
The overall radio spectra are determined by both shocks, when they coexist in a certain source. When BS's contribution is significant, it could form double-peaked or flat-top features in spectra (Figure \ref{fig2map}, left). Note that the flux displayed in the bottom left panel is calculated for $d_L=100$ Mpc. 
With {its} extraordinary high sensitivity, SKA will be powerful in discovering more radio afterglows at a flux level of $\simlt$50$\mu$Jy and characterizing the light curves and spectral evolution. These data will facilitate the studies on the origins of the radio emission, enable the identification of BS and clouds, and thereby reveal the circumnuclear environment in which the radio afterglows occur. 
{In the case of BS, we can use the radio flux and spectral evolution to map the
properties of gaseous clouds at a sub-pc scale \citep[e.g.,][]{yang2025}, which cannot be probed
via direct measurements even in the nearest supermassive black holes.
}

On the other hand, for nearby TDEs ($d_L < 100$ Mpc), high-resolution SKA+VLBI observations can distinguish the radio structure and help {to directly} identify the presence of a BS. We examined the case for an isotropic outflow with a velocity of 0.2 c impacting both a toroidal cloud (assuming to locate at 0.2 pc from the SMBH with a radius of 0.02 pc and a covering factor of 10\%) and a hot diffuse CNM with a Sgr A*-like density profile \citep{gillessen2019}. Figure \ref{fig2map} shows the synthetic radio images at various times and viewing angles (assuming at $d_L = 100$ Mpc and $d_A=96$ Mpc). The FS radio emission clearly expands over time, while the BS appears as stationary bright spots (in edge-on views) or a bright ring (in face-on views). Such imaging differences can also be used to determine the origin of the radio emission. 
Note that the radio flux density in our simulations from both bow shock and forward shock depends on the parameters for outflow, smooth CNM and dense clouds, and the actual flux can be lower than that shown in Figure \ref{fig2map}.  %, and the true flux can be lower by 
%SKA+VLBI observations 
{For most radio-emitting TDEs, with a spatial resolution of $\sim$0.5 mas at Band 5b, 
SKA+VLBI observations over a period of 3-5 years are able to resolve the radio emission of FS from BS (Figure \ref{fig2map}, right). 
}

%[ The high-resolution observations in the jet decelerating phase allow for studying how it affects the CNM environment. ]

\section{Unbiased detection of TDEs across diverse host galaxy environments}

%\subsection{Introduction }

%The host preference of post-starburst galaxies for optical TDEs, or just a selection effect?  The current status of selecting TDEs across different wavelengths, e.g., optical, infrared, X-ray and radio. 
%Specifically, SKA and VLBI observations of nuclear infrared flares are critical to identify dust-obscured TDEs, e.g., Arp 299-B AT1 \cite{Mattila2018}.  

%[LM Sun] 
Optical TDEs preferentially occur in post-starburst galaxies \citep{French2016}.
However, whether there are equally high TDE rates in star-forming and starburst galaxies remains unclear.
Although optical surveys rarely detect TDEs in starforming galaxies \citep[e.g.,][]{Yao2023}, this might be caused by dust obscuration \citep{Roth2021} and the intrinsic TDE rate may be high \citep{Tadhunter2017}.
To obtain an unbiased TDE demography, it is necessary to probe TDEs in the mid-infrared and radio bands with weak dust extinction.
Although some candidates of obscured TDEs have been discovered in these bands \citep{Jiang2021,Somalwar2022,Masterson2024}, their nature is still under debate.
The mid-infrared flare of Arp 299-B AT1 \citep{Mattila2018} %in an ultra-luminous infrared galaxy 
provides an example of how to probe the TDE nature through VLBI observations, 
which allow for detecting and measuring the jet proper motion (Figure \ref{fig1jet}).  
%It shows associated transient radio radiation, which was probed to originate from jet because VLBA monitoring measured proper motion.
The jet inclination is different from the axis of pre-existing accretion disk, indicating that the accreted material comes from high-inclination orbits, which strongly favors the TDE scenario. %suggests a TDE nature.
%In the future, SKA-based 
{Using  Arp 299-B AT1 as a prototype, long-term SKA+VLBI Band 5 observations of optically-dark infrared flares will help to identify more obscured TDEs by measuring the jet orientation and proper motion,} %thereby understanding 
uncovering the missing population of TDEs that occurred in starforming and starburst galaxies.

\subsection{Identification of off-nuclear TDEs}

%[TA] 
Off‑nuclear TDEs are expected in at least two contexts: (i) disruptions by IMBHs residing in nuclear star clusters, dwarf galaxies, or stripped nuclei; and (ii) disruptions by SMBHs that have been displaced from the galactic center by gravitational‑wave recoil or three‑body interactions during a merger. Both channels are already hinted at observationally {\citep[][under review]{Yao2025, Jin2025} }and they open a new window on black hole demographics and {studies of galaxy assembly.}  

% Identification of off-nuclear TDEs that could be linked to Intermediate-mass black holes or recoiling black hole \citep{Jin2025, Yao2024}?  The high sensitivity and spatial resolution is key to uncover the radio emission components, which makes sense for SKA-VLBI observations. 

%[SK] 
The unique combination of high sensitivity ($\mu$Jy) and spatial resolution (sub‑mas) offered by SKA+VLBI will allow a variety of applications related to off-nuclear TDEs. 
First, off-nuclear TDEs involving IMBHs are expected to form in nuclear star clusters.
%around intermediate-mass black holes in nuclear star clusters. 
Given the typical dimensions of nuclear starbursts of up to a few 100 pc \citep{Prieto2024}, these systems will be easily resolved with SKA+VLBI across a broad redshift range.  %out to redshifts of z=0.*.  
Further, dwarf galaxies with IMBHs frequently reside in galaxy halos. In that case, SKA+VLBI type resolution will be able to spatially resolve these systems out to much higher redshifts than is possible in the X-ray regime where some such systems have been detected so far \citep[][under review]{Lin2020, Li2026, Jin2025}, and thus will probe aspects of cosmic evolution as well. 
Second, the rate of TDEs is enhanced in the course of recoiling SMBHs that pass through the dense galactic environments \citep{KomossaMerritt2008}. Even if the galaxy's central region is obscured, radio observations will still penetrate the gas and dust and will provide an accurate measurement of the location of the recoiling SMBH within the nucleus \citep{Mattila2018, Yao2025}. Even if only a few radio jets from recoiling SMBHs can be identified, this will constitute a huge advancement of this new field of research. 

%[TA] 
Practical confirmation requires VLBI astrometry tied to the host nucleus. Phased SKA-Mid plus global VLBI deliver sub-mas imaging and $\lesssim$ tens-of-$\mu$as relative astrometry in $\sim1$\,h on-source time (rms of $\sim$3--10\,$\mu$Jy\,beam$^{-1}$), enabling to secure nuclear-offset measurements and multi-epoch tests for proper motion or expanding ejecta. Dual phase centres (transient + nucleus) mitigate registration systematics. In addition, high brightness temperatures ($> 10^{7}$~K), evolving synchrotron peaks, and polarimetry/rotation measure mapping are helpful to distinguish cluster/halo environments from nuclear screens. Triggers from Rubin/LSST, EP and SVOM will also provide the chance to capture early offsets and {centroid motion of transient radio emission from a newly launched jet}.

\subsection{Observations of radio emission from TDE candidates by supermassive black hole binaries}

%Can we use  SKA-VLBI to probe TDEs by supermassive black hole binary, e.g., jet precession? 

%[SK] 
Jetted TDEs are an excellent probe of binary SMBHs. In particular, they trace SMBH binaries even in quiescent galaxies without a permanent accretion disk. Such binary SMBHs are otherwise very challenging to detect, as most search methods require at least one SMBH to be active \citep{KomossaGrupe2024}. The lightcurve of a jetted TDE in a binary SMBH system will look characteristically different from a TDE in a single SMBH system \citep{LiuLi2009}: due to the influence of the secondary SMBH on the tidal stream towards the primary, accretion is temporarily interrupted and then restarts, leading to phases of characteristic dips and recoveries in the accretion lightcurve \citep{Liu2014, Shu2020}. These then affect the jet (if present) accordingly. 
That method alone is an important future SKA {application 
%(An et al. in this Book, and the binary-SMBH chapter). 
\citep[][in this Book]{TaoAn03.2026.SKA, DAmato01.2026.SKA}.}

%, not in itself requiring VLBI observations. However, if the binary SMBH was itself active and hosting a large-scale jet, then SKA-VLBI observations can additionally be used to trace the jet precession of the large-scale jet, providing unique novel tool of characterizing the binary SMBH orbital parameters. 

%[TA] 
In practice, phased SKA-Mid plus global VLBI can test the TDE scenario involving SMBHB through time-variable, frequency-dependent jet geometry and astrometry: (i) periodic swings in core position angle and {the electric vector polarization angle} %polarization EVPA 
(precession); (ii) quasi-periodic core-shift cycles across Bands 2--5b tied to modulated jet power; (iii) curved or oscillatory proper motions of compact knots; and (iv) rotation-measure sign changes expected from a precessing helical field. Monthly--quarterly monitoring over 1--3 yr with sub-mas imaging and $\lesssim$ tens-of-$\mu$as relative astrometry at $\mu$Jy sensitivities will separate binary-driven precession from single-BH Lense-Thirring disk precession and from stochastic variability, especially when combined with optical/X-ray dip--and--recovery light curves that flag stream interruptions. These capabilities follow from the SKA+VLBI performance and band coverage summarized for AA$^*$/AA4. 
On the other hand, if the binary SMBH was itself active and hosting a large-scale jet, then SKA+VLBI observations can additionally be used to trace the jet precession of the large-scale jet, providing unique novel tool of characterizing the binary SMBH orbital parameters.

\section{Uncovering the physics of the TDE-neutrinos association}
 %  [The detection of neutrinos in TDEs \citep{Stein2021} and possible models involving jets/outflows.    Localizing the radio counterpart of high-energy neutrinos and identifying the temporal coincidence with high-cadence SKA survey observations. By tracing the evolution of the radio structure or proper motion, the SKA-VLBI observations enable us to establish the connection between radio flares and neutrinos, and disentangle the critical role of jet/outflow in producing neutrinos, providing important constraints on theoretical models. ]

%[SK] 
%TDEs as multimessenger sources: 
TDEs are inherently multiwavelength and multimessenger sources \citep[e.g.,][]{Komossa2015}. 
In a variety of situations, the radio regime provides unique information that is either linked to questions of jet formation and evolution, and/or to the high-precision spatial localization of the TDE, or even TDE identification in heavily obscured environments where optical and X-ray detection methods fail. In addition, high-sensitivity radio observations are an important ingredient in the multi-messenger approach: jetted TDEs are candidates for neutrino emitters  \citep{Stein2021, Mohan2022}, and they trace a parameter space very different from blazars: the majority of TDEs are identified in quiescent host galaxies, implying a pristine environment without any previous large- or small-scale jet activity. Further, the presence and properties/configuration of magnetic fields are expected to be different in TDEs that form a compact accretion disk within a short amount of time. 
It is also suggested that the efficiency of high-energy neutrino production in TDEs could be increased compared to non-flaring {active galactic nuclei (AGN)}, possibly due to the high Eddington ratio of the TDE flares \citep{vanVelzen2024}.
%Further, TDEs are also sources of gravitational waves. An extreme mass ratio event \citep{Amaro-Seoane2018}, like a white dwarf partially disrupted by a SMBH, will be source of both, gravitational wave (GW) and electromagnetic radiation. While the GW emission (in the LISA sensitivity band) will have very large astrometric uncertainties SKA+VLBI will allow a high-precision spatial location of any jetted event. 
These aspects make neutrino and radio observations of TDE jets important tracers of jet physics in a unique parameter space. 
On the other hand, if a TDE occurs in a gas-rich environment with dense clouds at a sub-pc scale (Section 4.2), the outflow-cloud interactions may form shock waves which generate accelerated protons, and hence delayed neutrinos from hadronic interactions in clouds \citep{Wu2022, Zhou2026}. 
However, the predicted neutrino number in such a scenario depends on various parameters such as the density and size of the clouds, the covering factor of BS, and shock acceleration efficiency, which cannot be measured directly, even for the nearest TDEs, but could be constrained with radio observations. 
{As suggested by \citet{Zhou2026}, the scenario of outflow-cloud interaction in producing the neutrino emission through pp collisions can be tested with future
identifications of radio transients coincident with high-energy neutrinos.
}

%in order to match the observed sub-PeV neutrino event possibly associated with the TDE AT2019dsg \citep{Stein2021}, 

%[TA] 
Therefore, beyond electromagnetic diagnostics, SKA+VLBI can critically test the emerging TDE--neutrino connection. Hadronic models predict high-energy neutrinos from p$\gamma$/pp interactions in jetted or shock-powered TDE outflows, with peak production contemporaneous with the radio-bright phase. Sub-mas imaging and $\lesssim$tens-of-$\mu$as (relative) astrometry will localize the synchrotron zone with respect to the nucleus, measure core-shift magnetic fields, and provide kinetic-energy and density calorimetry needed to infer neutrino efficiencies. Rapid targets of opportunity observations to IceCube/KM3NeT/Baikal-GVD EHE/Gold alerts, plus stacking of radio-selected TDE samples, can confirm or refute event-level associations (e.g., AT2019dsg, AT2019fdr and SDSS J1513+3111) and distinguish hadronic from leptonic power. A coordinated program (Bands 2–5b, monthly cadence) will set population-level limits on the TDE contribution to the diffuse neutrino background (e.g., \citealt{Murase2016, Stein2021, Reusch2022}).
 
   \section{ SKA-VLBI observations of transient jet/outflows from other types of SMBH accretion flares } 

   %The results from the VLBI observations of the ``changing-look" AGN 1ES 1927+654 \citep{Meyer2025}. 

   In addition to TDEs, 
%modern time-domainspectral observations have revealed 
another recent advance in time-domain studies is the identifications of {non-regular} extreme variability in active
galaxies, which can change their
optical types on timescales of years, characterized by emerged
or disappeared broad emission lines as well as dramatic
continuum variations \citep[e.g.,][]{Shappee2014}. 
This class of ``changing-look” AGN (CLAGNs), 
may even be transitioning between radiatively inefficient and radiatively efficient modes of accretion \citep{Ruan2019}, analogous to X-ray spectral state transitions observed in stellar-mass BHs. 
{On the other hand, there is a new class of flares from accreting SMBHs in AGNs that decline more slowly than normal TDEs, perhaps arising from a longer-term event of intensified gas accretion \citep{Trakhtenbrot2019b}. 
Another class of SMBH-related accreting flares exhibit characteristics of both AGNs and TDEs, which cannot easily be classified into either source class and have been dubbed as ambiguous nuclear transients \citep{Neustadt2020, Hinkle2022}. 
 Radio observations of these SMBH-related accreting flares}
 are important to disclose the connection between SMBH accretion and the launching of jets and outflows in a possibly different parameter space from TDEs.  

 The AGN 1ES 1927+654 is one of few extreme CLAGNs
observed to change in the black hole's accretion states in real time \citep{Trakhtenbrot2019a}, 
{which could be triggered by the interaction between the preexisting accretion
disk and the fallback debris from TDE \citep{Ricci2020}. }
%Over five years after observing it being a CLAGN, 
Approximately 1800 days after the initial CL event, a significant radio brightening was observed in 1ES 1927+654, consistent with a newly launched radio-emitting outflow. 
At the resolution of $\sim$0.5 mas, VLBA imaging shows 
bipolar radio extensions of similar brightness separated by
approximately 0.45 mas or 0.15 pc, with a tentative expansion
of separation speed of $\sim$0.3c \citep{Meyer2025}. 
{Future wide-field SKA slow-transient surveys will be sensitive enough to capture the transient radio brightening following the AGN flares. 
We advocate a SKA+VLBI program at Band 5 to track a sample of 10-20 extreme AGN flares similar to 1ES 1927+654, with 3-4 epochs per source spanning a few weeks to years after the discovery, 
%Future radio monitoring observations of {\bf extreme AGN flares similar to 1ES 1927+654} are encouraged to 
which are necessary to catch the newly launched radio-emitting outflow or monitor its temporal evolution. 
With improved sensitivity and spatial resolution, SKA+VLBI will enable to detect the resolved jet structures for AGN flares at a distance of $\simlt$100 Mpc, allowing for more detailed studies on 
%to further constrain 
the kinematics and energetics of the jet in its earliest evolution phases. } 
%in this ever-changing and unique AGN. 

\bigskip
{\bf Acknowledgments:} 
We thank the anonymous reviewer for detailed and insightful comments that helped to improve the chapter 
significantly.  
LWH acknowledges
support from the National Natural Science Foundation of China (grant Nos. 12473012). MPT acknowledges financial support from the Severo Ochoa grant
CEX2021-001131-S and from the Spanish grant PID2023-147883NB-C21, funded by MCIU/AEI/ 10.13039/501100011033, as well as support through ERDF/EU.
X.S. acknowledges support from the National Science Foundation of China through grant No. 12192220 and 12192221.

\bibliographystyle{abbrvnat-maxbibnames4}
\bibliography{chapter} % if your bibtex file is called example.bib

@ARTICLE{Li2026,
        author = {{Li}, D. -Y. and {Yang}, J. and {Zhang}, W. -D. and {the Einstein Probe collaborations}},
        title = "{A fast powerful X-ray transient from possible tidal disruption of a white dwarf}",
      journal = {Science Bulletin},
         year = 2026,
        month = feb,
       volume = {71},
       number = {3},
        pages = {538-546},
          doi = {10.1016/j.scib.2025.12.050},
archivePrefix = {arXiv},
       eprint = {2509.25877},
 primaryClass = {astro-ph.HE},
       adsurl = {https://ui.adsabs.harvard.edu/abs/2026SciBu..71..538L}
}

@ARTICLE{Trakhtenbrot2019a,
       author = {{Trakhtenbrot}, Benny and {Arcavi}, Iair and {Ricci}, Claudio and {Tacchella}, Sandro and {Stern}, Daniel and {Netzer}, Hagai and {Jonker}, Peter G. and {Horesh}, Assaf and {Mej{\'\i}a-Restrepo}, Juli{\'a}n Esteban and {Hosseinzadeh}, Griffin and {Hallefors}, Valentina and {Howell}, D. Andrew and {McCully}, Curtis and {Balokovi{\'c}}, Mislav and {Heida}, Marianne and {Kamraj}, Nikita and {Lansbury}, George Benjamin and {Wyrzykowski}, {\L}ukasz and {Gromadzki}, Mariusz and {Hamanowicz}, Aleksandra and {Cenko}, S. Bradley and {Sand}, David J. and {Hsiao}, Eric Y. and {Phillips}, Mark M. and {Diamond}, Tiara R. and {Kara}, Erin and {Gendreau}, Keith C. and {Arzoumanian}, Zaven and {Remillard}, Ron},
        title = "{A new class of flares from accreting supermassive black holes}",
      journal = {Nature Astronomy},
         year = 2019,
        month = jan,
       volume = {3},
        pages = {242-250},
          doi = {10.1038/s41550-018-0661-3},
archivePrefix = {arXiv},
       eprint = {1901.03731},
 primaryClass = {astro-ph.GA},
       adsurl = {https://ui.adsabs.harvard.edu/abs/2019NatAs...3..242T}
}

@ARTICLE{Hinkle2022,
       author = {{Hinkle}, Jason T. and {Holoien}, Thomas W. -S. and {Shappee}, Benjamin. J. and {Neustadt}, Jack M.~M. and {Auchettl}, Katie and {Vallely}, Patrick J. and {Shahbandeh}, Melissa and {Kluge}, Matthias and {Kochanek}, Christopher S. and {Stanek}, K.~Z. and {Huber}, Mark E. and {Post}, Richard S. and {Bersier}, David and {Ashall}, Christopher and {Tucker}, Michael A. and {Williams}, Jonathan P. and {de Jaeger}, Thomas and {Do}, Aaron and {Fausnaugh}, Michael and {Gruen}, Daniel and {Hopp}, Ulrich and {Myles}, Justin and {Obermeier}, Christian and {Payne}, Anna V. and {Thompson}, Todd A.},
        title = "{The Curious Case of ASASSN-20hx: A Slowly Evolving, UV- and X-Ray-Luminous, Ambiguous Nuclear Transient}",
      journal = {\apj},
         year = 2022,
        month = may,
       volume = {930},
       number = {1},
          eid = {12},
        pages = {12},
          doi = {10.3847/1538-4357/ac5f54},
archivePrefix = {arXiv},
       eprint = {2108.03245},
 primaryClass = {astro-ph.HE},
       adsurl = {https://ui.adsabs.harvard.edu/abs/2022ApJ...930...12H}
}

@ARTICLE{Neustadt2020,
       author = {{Neustadt}, J.~M.~M. and {Holoien}, T.~W. -S. and {Kochanek}, C.~S. and {Auchettl}, K. and {Brown}, J.~S. and {Shappee}, B.~J. and {Pogge}, R.~W. and {Dong}, Subo and {Stanek}, K.~Z. and {Tucker}, M.~A. and {Bose}, S. and {Chen}, Ping and {Ricci}, C. and {Vallely}, P.~J. and {Prieto}, J.~L. and {Thompson}, T.~A. and {Coulter}, D.~A. and {Drout}, M.~R. and {Foley}, R.~J. and {Kilpatrick}, C.~D. and {Piro}, A.~L. and {Rojas-Bravo}, C. and {Buckley}, D.~A.~H. and {Gromadzki}, M. and {Dimitriadis}, G. and {Siebert}, M.~R. and {Do}, A. and {Huber}, M.~E. and {Payne}, A.~V.},
        title = "{To TDE or not to TDE: the luminous transient ASASSN-18jd with TDE-like and AGN-like qualities}",
      journal = {\mnras},
         year = 2020,
        month = may,
       volume = {494},
       number = {2},
        pages = {2538-2560},
          doi = {10.1093/mnras/staa859},
archivePrefix = {arXiv},
       eprint = {1910.01142},
 primaryClass = {astro-ph.HE},
       adsurl = {https://ui.adsabs.harvard.edu/abs/2020MNRAS.494.2538N}
}

@ARTICLE{Zhou2026,
       author = {{Zhou}, Tianyao and {Shu}, Xinwen and {Mou}, Guobin and {Yang}, Lei and {Sun}, Luming and {Peng}, Fangkun and {Zhang}, Fabao and {Ding}, Hucheng and {Jiang}, Ning and {Wang}, Tinggui and {Chandola}, Yogesh and {Liu}, Daizhong and {Dou}, Liming and {Wang}, Yibo and {Wang}, Jianguo and {Wu}, Zhongzu and {Yang}, Chenwei},
        title = "{Dust-obscured radio-emitting tidal disruption event coincident with a high-energy neutrino event}",
      journal = {\prd},
         year = 2026,
        month = feb,
       volume = {113},
       number = {4},
          eid = {043046},
        pages = {043046},
          doi = {10.1103/j8g9-f6hh},
archivePrefix = {arXiv},
       eprint = {2601.05601},
 primaryClass = {astro-ph.HE},
       adsurl = {https://ui.adsabs.harvard.edu/abs/2026PhRvD.113d3046Z}
}

@ARTICLE{Brown2015,
       author = {{Brown}, G.~C. and {Levan}, A.~J. and {Stanway}, E.~R. and {Tanvir}, N.~R. and {Cenko}, S.~B. and {Berger}, E. and {Chornock}, R. and {Cucchiaria}, A.},
        title = "{Swift J1112.2-8238: a candidate relativistic tidal disruption flare}",
      journal = {\mnras},
         year = 2015,
        month = oct,
       volume = {452},
       number = {4},
        pages = {4297-4306},
          doi = {10.1093/mnras/stv1520},
archivePrefix = {arXiv},
       eprint = {1507.03582},
 primaryClass = {astro-ph.HE},
       adsurl = {https://ui.adsabs.harvard.edu/abs/2015MNRAS.452.4297B}
}

@ARTICLE{Cenko2012,
       author = {{Cenko}, S. Bradley and {Krimm}, Hans A. and {Horesh}, Assaf and {Rau}, Arne and {Frail}, Dale A. and {Kennea}, Jamie A. and {Levan}, Andrew J. and {Holland}, Stephen T. and {Butler}, Nathaniel R. and {Quimby}, Robert M. and {Bloom}, Joshua S. and {Filippenko}, Alexei V. and {Gal-Yam}, Avishay and {Greiner}, Jochen and {Kulkarni}, S.~R. and {Ofek}, Eran O. and {Olivares E.}, Felipe and {Schady}, Patricia and {Silverman}, Jeffrey M. and {Tanvir}, Nial R. and {Xu}, Dong},
        title = "{Swift J2058.4+0516: Discovery of a Possible Second Relativistic Tidal Disruption Flare?}",
      journal = {\apj},
         year = 2012,
        month = jul,
       volume = {753},
       number = {1},
          eid = {77},
        pages = {77},
          doi = {10.1088/0004-637X/753/1/77},
archivePrefix = {arXiv},
       eprint = {1107.5307},
 primaryClass = {astro-ph.HE},
       adsurl = {https://ui.adsabs.harvard.edu/abs/2012ApJ...753...77C}
}

@ARTICLE{Matsumoto2023,
       author = {{Matsumoto}, Tatsuya and {Piran}, Tsvi},
        title = "{Generalized equipartition method from an arbitrary viewing angle}",
      journal = {\mnras},
         year = 2023,
        month = jul,
       volume = {522},
       number = {3},
        pages = {4565-4576},
          doi = {10.1093/mnras/stad1269},
archivePrefix = {arXiv},
       eprint = {2211.10051},
 primaryClass = {astro-ph.HE},
       adsurl = {https://ui.adsabs.harvard.edu/abs/2023MNRAS.522.4565M}
}

@ARTICLE{Matsumoto2024,
       author = {{Matsumoto}, Tatsuya and {Piran}, Tsvi},
        title = "{Late-time Radio Flares in Tidal Disruption Events}",
      journal = {\apj},
         year = 2024,
        month = aug,
       volume = {971},
       number = {1},
          eid = {49},
        pages = {49},
          doi = {10.3847/1538-4357/ad58ba},
archivePrefix = {arXiv},
       eprint = {2404.15966},
 primaryClass = {astro-ph.HE},
       adsurl = {https://ui.adsabs.harvard.edu/abs/2024ApJ...971...49M}
}

@ARTICLE{Lu2024,
       author = {{Lu}, Wenbin and {Matsumoto}, Tatsuya and {Matzner}, Christopher D.},
        title = "{Misaligned precessing jets are choked by the accretion disc wind}",
      journal = {\mnras},
         year = 2024,
        month = sep,
       volume = {533},
       number = {1},
        pages = {979-993},
          doi = {10.1093/mnras/stae1770},
archivePrefix = {arXiv},
       eprint = {2310.15336},
 primaryClass = {astro-ph.HE},
       adsurl = {https://ui.adsabs.harvard.edu/abs/2024MNRAS.533..979L}
}

@ARTICLE{Teboul2023,
       author = {{Teboul}, Odelia and {Metzger}, Brian D.},
        title = "{A Unified Theory of Jetted Tidal Disruption Events: From Promptly Escaping Relativistic to Delayed Transrelativistic Jets}",
      journal = {\apjl},
         year = 2023,
        month = nov,
       volume = {957},
       number = {1},
          eid = {L9},
        pages = {L9},
          doi = {10.3847/2041-8213/ad0037},
archivePrefix = {arXiv},
       eprint = {2308.05161},
 primaryClass = {astro-ph.HE},
       adsurl = {https://ui.adsabs.harvard.edu/abs/2023ApJ...957L...9T}
}

@ARTICLE{Cendes2024,
       author = {{Cendes}, Y. and {Berger}, E. and {Alexander}, K.~D. and {Chornock}, R. and {Margutti}, R. and {Metzger}, B. and {Wieringa}, M.~H. and {Bietenholz}, M.~F. and {Hajela}, A. and {Laskar}, T. and {Stroh}, M.~C. and {Terreran}, G.},
        title = "{Ubiquitous Late Radio Emission from Tidal Disruption Events}",
      journal = {\apj},
         year = 2024,
        month = aug,
       volume = {971},
       number = {2},
          eid = {185},
        pages = {185},
          doi = {10.3847/1538-4357/ad5541},
archivePrefix = {arXiv},
       eprint = {2308.13595},
 primaryClass = {astro-ph.HE},
       adsurl = {https://ui.adsabs.harvard.edu/abs/2024ApJ...971..185C}
}

@ARTICLE{Cendes2022,
       author = {{Cendes}, Y. and {Berger}, E. and {Alexander}, K.~D. and {Gomez}, S. and {Hajela}, A. and {Chornock}, R. and {Laskar}, T. and {Margutti}, R. and {Metzger}, B. and {Bietenholz}, M.~F. and {Brethauer}, D. and {Wieringa}, M.~H.},
        title = "{A Mildly Relativistic Outflow Launched Two Years after Disruption in Tidal Disruption Event AT2018hyz}",
      journal = {\apj},
         year = 2022,
        month = oct,
       volume = {938},
       number = {1},
          eid = {28},
        pages = {28},
          doi = {10.3847/1538-4357/ac88d0},
archivePrefix = {arXiv},
       eprint = {2206.14297},
 primaryClass = {astro-ph.HE},
       adsurl = {https://ui.adsabs.harvard.edu/abs/2022ApJ...938...28C}
}

@ARTICLE{Burrows2011,
       author = {{Burrows}, D.~N. and {Kennea}, J.~A. and {Ghisellini}, G. and {Mangano}, V. and {Zhang}, B. and {Page}, K.~L. and {Eracleous}, M. and {Romano}, P. and {Sakamoto}, T. and {Falcone}, A.~D. and {Osborne}, J.~P. and {Campana}, S. and {Beardmore}, A.~P. and {Breeveld}, A.~A. and {Chester}, M.~M. and {Corbet}, R. and {Covino}, S. and {Cummings}, J.~R. and {D'Avanzo}, P. and {D'Elia}, V. and {Esposito}, P. and {Evans}, P.~A. and {Fugazza}, D. and {Gelbord}, J.~M. and {Hiroi}, K. and {Holland}, S.~T. and {Huang}, K.~Y. and {Im}, M. and {Israel}, G. and {Jeon}, Y. and {Jeon}, Y.-B. and {Jun}, H.~D. and {Kawai}, N. and {Kim}, J.~H. and {Krimm}, H.~A. and {Marshall}, F.~E. and {P. M{\'e}sz{\'a}ros} and {Negoro}, H. and {Omodei}, N. and {Park}, W.-K. and {Perkins}, J.~S. and {Sugizaki}, M. and {Sung}, H.-I. and {Tagliaferri}, G. and {Troja}, E. and {Ueda}, Y. and {Urata}, Y. and {Usui}, R. and {Antonelli}, L.~A. and {Barthelmy}, S.~D. and {Cusumano}, G. and {Giommi}, P. and {Melandri}, A. and {Perri}, M. and {Racusin}, J.~L. and {Sbarufatti}, B. and {Siegel}, M.~H. and {Gehrels}, N.},
        title = "{Relativistic jet activity from the tidal disruption of a star by a massive black hole}",
      journal = {\nat},
         year = 2011,
        month = aug,
       volume = {476},
       number = {7361},
        pages = {421-424},
          doi = {10.1038/nature10374},
archivePrefix = {arXiv},
       eprint = {1104.4787},
 primaryClass = {astro-ph.HE},
       adsurl = {https://ui.adsabs.harvard.edu/abs/2011Natur.476..421B}
}

@ARTICLE{Shu2025,
       author = {{Shu}, Xinwen and {Yang}, Lei and {Yang}, Haonan and {Xu}, Fan and {Chen}, Jin-Hong and {Eyles-Ferris}, Rob A.~J. and {Dai}, Lixin and {Yu}, Yunwei and {Shen}, Rong-Feng and {Sun}, Luming and {Ding}, Hucheng and {Zheng}, WeiKang and {Jiang}, Ning and {Li}, Wenxiong and {Sun}, Ning-Chen and {Xu}, Dong and {Zhang}, Zhumao and {Jin}, Chichuan and {Rau}, Arne and {Wang}, Tinggui and {Wu}, Xue-feng and {Yuan}, Weimin and {Zhang}, Bing and {Nandra}, Kirpal and {Filippenko}, Alexei V. and {Poidevin}, Fr{\'e}d{\'e}rick and {Soria}, Roberto and {Kumar}, Amit and {Aguado}, David S. and {An}, Fangxia and {An}, Tao and {An}, Jie and {Andrews}, Moira and {Anutarawiramkul}, Rungrit and {Baldini}, Pietro and {Brink}, Thomas G. and {Butpan}, Pathompong and {Cai}, Zhiming and {Castro-Tirado}, Alberto J. and {Cheng}, Huaqing and {Cui}, Weiwei and {Farah}, Joseph and {Fu}, Shaoyu and {Fynbo}, Johan P.~U. and {Gao}, Xing and {Han}, Dawei and {Han}, Xuhui and {Howell}, D. Andrew and {Hu}, Jingwei and {Jiang}, Shuaiqing and {Kumar}, Brajesh and {Lei}, Weihua and {Li}, Dongyue and {Li}, Chengkui and {Liu}, Huaqiu and {Liu}, Xing and {Liu}, Yuan and {Liu}, Xiaowei and {L{\'o}pez-Oramas}, Alicia and {L{\'o}pez Fern{\'a}ndez-Nespral}, David and {Maund}, Justyn R. and {McCully}, Curtis and {Niu}, Zexi and {Newsome}, Megan and {O'Brien}, Paul and {Pan}, Haiwu and {Pan}, Yu and {Padilla Gonzalez}, Estefania and {P{\'e}rez-Fournon}, Ismael and {Silima}, Walter and {Sun}, Hui and {Sun}, Shengli and {Sun}, Xiaojin and {Terreran}, Giacomo and {Tinyanont}, Samaporn and {Wang}, Junxian and {Wang}, Yanan and {Wang}, Yun and {Wiersema}, Klaas and {Xu}, Yunfei and {Xue}, Yongquan and {Yang}, Yi and {Zhang}, Fabao and {Zhang}, Juan and {Zhang}, Pinpin and {Zhang}, Wenda and {Zhang}, Yonghe and {Zhao}, Haisheng and {Zhu}, Zipei and {Xin}, Liping and {Yao}, Zhuheng and {Cordier}, Bertrand and {Wei}, Jianyan and {Qiu}, Yulei and {Daigne}, Fr{\'e}d{\'e}ric},
        title = "{EP241021a: A Months-duration X-Ray Transient with Luminous Optical and Radio Emission}",
      journal = {\apjl},
         year = 2025,
        month = sep,
       volume = {990},
       number = {1},
          eid = {L29},
        pages = {L29},
          doi = {10.3847/2041-8213/adf4cd},
archivePrefix = {arXiv},
       eprint = {2505.07665},
 primaryClass = {astro-ph.HE},
       adsurl = {https://ui.adsabs.harvard.edu/abs/2025ApJ...990L..29S}
}

@ARTICLE{gillessen2019,
       author = {{Gillessen}, S. and {Plewa}, P.~M. and {Widmann}, F. and {von Fellenberg}, S. and {Schartmann}, M. and {Habibi}, M. and {Jimenez Rosales}, A. and {Baub{\"o}ck}, M. and {Dexter}, J. and {Gao}, F. and {Waisberg}, I. and {Eisenhauer}, F. and {Pfuhl}, O. and {Ott}, T. and {Burkert}, A. and {de Zeeuw}, P.~T. and {Genzel}, R.},
        title = "{Detection of a Drag Force in G2's Orbit: Measuring the Density of the Accretion Flow onto Sgr A* at 1000 Schwarzschild Radii}",
      journal = {\apj},
         year = 2019,
        month = jan,
       volume = {871},
       number = {1},
          eid = {126},
        pages = {126},
          doi = {10.3847/1538-4357/aaf4f8},
       adsurl = {https://ui.adsabs.harvard.edu/abs/2019ApJ...871..126G}
}

@ARTICLE{Horesh2021,
       author = {{Horesh}, A. and {Cenko}, S.~B. and {Arcavi}, I.},
        title = "{Delayed radio flares from a tidal disruption event}",
      journal = {Nature Astronomy},
         year = 2021,
        month = may,
       volume = {5},
        pages = {491-497},
          doi = {10.1038/s41550-021-01300-8},
archivePrefix = {arXiv},
       eprint = {2102.11290},
 primaryClass = {astro-ph.HE},
       adsurl = {https://ui.adsabs.harvard.edu/abs/2021NatAs...5..491H}
}

@ARTICLE{Tchekhovskoy2014,
       author = {{Tchekhovskoy}, Alexander and {Metzger}, Brian D. and {Giannios}, Dimitrios and {Kelley}, Luke Z.},
        title = "{Swift J1644+57 gone MAD: the case for dynamically important magnetic flux threading the black hole in a jetted tidal disruption event}",
      journal = {\mnras},
         year = 2014,
        month = jan,
       volume = {437},
       number = {3},
        pages = {2744-2760},
          doi = {10.1093/mnras/stt2085},
archivePrefix = {arXiv},
       eprint = {1301.1982},
 primaryClass = {astro-ph.HE},
       adsurl = {https://ui.adsabs.harvard.edu/abs/2014MNRAS.437.2744T}
}

@ARTICLE{Komossa2015,
       author = {{Komossa}, S.},
        title = "{Tidal disruption of stars by supermassive black holes: Status of observations}",
      journal = {Journal of High Energy Astrophysics},
         year = 2015,
        month = sep,
       volume = {7},
        pages = {148-157},
          doi = {10.1016/j.jheap.2015.04.006},
archivePrefix = {arXiv},
       eprint = {1505.01093},
 primaryClass = {astro-ph.HE},
       adsurl = {https://ui.adsabs.harvard.edu/abs/2015JHEAp...7..148K}
}

@ARTICLE{Rees1988,
       author = {{Rees}, Martin J.},
        title = "{Tidal disruption of stars by black holes of {}10$^{6}$-{}10$^{8}$ solar masses in nearby galaxies}",
      journal = {\nat},
         year = 1988,
        month = jun,
       volume = {333},
       number = {6173},
        pages = {523-528},
          doi = {10.1038/333523a0},
       adsurl = {https://ui.adsabs.harvard.edu/abs/1988Natur.333..523R}
}

@ARTICLE{Ruan2019,
       author = {{Ruan}, John J. and {Anderson}, Scott F. and {Eracleous}, Michael and {Green}, Paul J. and {Haggard}, Daryl and {MacLeod}, Chelsea L. and {Runnoe}, Jessie C. and {Sobolewska}, Malgosia A.},
        title = "{The Analogous Structure of Accretion Flows in Supermassive and Stellar Mass Black Holes: New Insights from Faded Changing-look Quasars}",
      journal = {\apj},
         year = 2019,
        month = sep,
       volume = {883},
       number = {1},
          eid = {76},
        pages = {76},
          doi = {10.3847/1538-4357/ab3c1a},
archivePrefix = {arXiv},
       eprint = {1903.02553},
 primaryClass = {astro-ph.HE},
       adsurl = {https://ui.adsabs.harvard.edu/abs/2019ApJ...883...76R}
}

@ARTICLE{Shappee2014,
       author = {{Shappee}, B.~J. and {Prieto}, J.~L. and {Grupe}, D. and {Kochanek}, C.~S. and {Stanek}, K.~Z. and {De Rosa}, G. and {Mathur}, S. and {Zu}, Y. and {Peterson}, B.~M. and {Pogge}, R.~W. and {Komossa}, S. and {Im}, M. and {Jencson}, J. and {Holoien}, T.~W.-S. and {Basu}, U. and {Beacom}, J.~F. and {Szczygie{\l}}, D.~M. and {Brimacombe}, J. and {Adams}, S. and {Campillay}, A. and {Choi}, C. and {Contreras}, C. and {Dietrich}, M. and {Dubberley}, M. and {Elphick}, M. and {Foale}, S. and {Giustini}, M. and {Gonzalez}, C. and {Hawkins}, E. and {Howell}, D.~A. and {Hsiao}, E.~Y. and {Koss}, M. and {Leighly}, K.~M. and {Morrell}, N. and {Mudd}, D. and {Mullins}, D. and {Nugent}, J.~M. and {Parrent}, J. and {Phillips}, M.~M. and {Pojmanski}, G. and {Rosing}, W. and {Ross}, R. and {Sand}, D. and {Terndrup}, D.~M. and {Valenti}, S. and {Walker}, Z. and {Yoon}, Y.},
        title = "{The Man behind the Curtain: X-Rays Drive the UV through NIR Variability in the 2013 Active Galactic Nucleus Outburst in NGC 2617}",
      journal = {\apj},
         year = 2014,
        month = jun,
       volume = {788},
       number = {1},
          eid = {48},
        pages = {48},
          doi = {10.1088/0004-637X/788/1/48},
archivePrefix = {arXiv},
       eprint = {1310.2241},
 primaryClass = {astro-ph.HE},
       adsurl = {https://ui.adsabs.harvard.edu/abs/2014ApJ...788...48S}
}

@ARTICLE{Baganoff03,
       author = {{Baganoff}, F.~K. and {Maeda}, Y. and {Morris}, M. and {Bautz}, M.~W. and {Brandt}, W.~N. and {Cui}, W. and {Doty}, J.~P. and {Feigelson}, E.~D. and {Garmire}, G.~P. and {Pravdo}, S.~H. and {Ricker}, G.~R. and {Townsley}, L.~K.},
        title = "{Chandra X-Ray Spectroscopic Imaging of Sagittarius A* and the Central Parsec of the Galaxy}",
      journal = {\apj},
         year = 2003,
        month = jul,
       volume = {591},
       number = {2},
        pages = {891-915},
          doi = {10.1086/375145},
archivePrefix = {arXiv},
       eprint = {astro-ph/0102151},
 primaryClass = {astro-ph},
       adsurl = {https://ui.adsabs.harvard.edu/abs/2003ApJ...591..891B}
}

@ARTICLE{Anderson20,
       author = {{Anderson}, M.~M. and {Mooley}, K.~P. and {Hallinan}, G. and {Dong}, D. and {Phinney}, E.~S. and {Horesh}, A. and {Bourke}, S. and {Cenko}, S.~B. and {Frail}, D. and {Kulkarni}, S.~R. and {Myers}, S.},
        title = "{Caltech-NRAO Stripe 82 Survey (CNSS). III. The First Radio-discovered Tidal Disruption Event, CNSS J0019+00}",
      journal = {\apj},
         year = 2020,
        month = nov,
       volume = {903},
       number = {2},
          eid = {116},
        pages = {116},
          doi = {10.3847/1538-4357/abb94b},
archivePrefix = {arXiv},
       eprint = {1910.11912},
 primaryClass = {astro-ph.HE},
       adsurl = {https://ui.adsabs.harvard.edu/abs/2020ApJ...903..116A}
}

@ARTICLE{Mattila2018,
       author = {{Mattila}, S. and {P{\'e}rez-Torres}, M. and {Efstathiou}, A. and {Mimica}, P. and {Fraser}, M. and {Kankare}, E. and {Alberdi}, A. and {Aloy}, M. {\'A}. and {Heikkil{\"a}}, T. and {Jonker}, P.~G. and {Lundqvist}, P. and {Mart{\'\i}-Vidal}, I. and {Meikle}, W.~P.~S. and {Romero-Ca{\~n}izales}, C. and {Smartt}, S.~J. and {Tsygankov}, S. and {Varenius}, E. and {Alonso-Herrero}, A. and {Bondi}, M. and {Fransson}, C. and {Herrero-Illana}, R. and {Kangas}, T. and {Kotak}, R. and {Ram{\'\i}rez-Olivencia}, N. and {V{\"a}is{\"a}nen}, P. and {Beswick}, R.~J. and {Clements}, D.~L. and {Greimel}, R. and {Harmanen}, J. and {Kotilainen}, J. and {Nandra}, K. and {Reynolds}, T. and {Ryder}, S. and {Walton}, N.~A. and {Wiik}, K. and {{\"O}stlin}, G.},
        title = "{A dust-enshrouded tidal disruption event with a resolved radio jet in a galaxy merger}",
      journal = {Science},
         year = 2018,
        month = aug,
       volume = {361},
       number = {6401},
        pages = {482-485},
          doi = {10.1126/science.aao4669},
archivePrefix = {arXiv},
       eprint = {1806.05717},
 primaryClass = {astro-ph.GA},
       adsurl = {https://ui.adsabs.harvard.edu/abs/2018Sci...361..482M}
}

@ARTICLE{Eftekhari18,
       author = {{Eftekhari}, T. and {Berger}, E. and {Zauderer}, B.~A. and {Margutti}, R. and {Alexander}, K.~D.},
        title = "{Radio Monitoring of the Tidal Disruption Event Swift J164449.3+573451. III. Late-time Jet Energetics and a Deviation from Equipartition}",
      journal = {\apj},
         year = 2018,
        month = feb,
       volume = {854},
       number = {2},
          eid = {86},
        pages = {86},
          doi = {10.3847/1538-4357/aaa8e0},
archivePrefix = {arXiv},
       eprint = {1710.07289},
 primaryClass = {astro-ph.HE},
       adsurl = {https://ui.adsabs.harvard.edu/abs/2018ApJ...854...86E}
}

@ARTICLE{Coughlin2014,
       author = {{Coughlin}, Eric R. and {Begelman}, Mitchell C.},
        title = "{Hyperaccretion during Tidal Disruption Events: Weakly Bound Debris Envelopes and Jets}",
      journal = {\apj},
         year = 2014,
        month = feb,
       volume = {781},
       number = {2},
          eid = {82},
        pages = {82},
          doi = {10.1088/0004-637X/781/2/82},
archivePrefix = {arXiv},
       eprint = {1312.5314},
 primaryClass = {astro-ph.HE},
       adsurl = {https://ui.adsabs.harvard.edu/abs/2014ApJ...781...82C}
}

@ARTICLE{Zhou2024,
       author = {{Zhou}, Chang and {Zhu}, Zi-Pei and {Lei}, Wei-Hua and {Fu}, Shao-Yu and {Xie}, Wei and {Xu}, Dong},
        title = "{AT2022cmc: A Tidal Disruption Event with a Two-component Jet in a Bondi-profile Circumnuclear Medium}",
      journal = {\apj},
         year = 2024,
        month = mar,
       volume = {963},
       number = {1},
          eid = {66},
        pages = {66},
          doi = {10.3847/1538-4357/ad20f3},
       adsurl = {https://ui.adsabs.harvard.edu/abs/2024ApJ...963...66Z}
}

@ARTICLE{Pasham2018,
       author = {{Pasham}, Dheeraj R. and {van Velzen}, Sjoert},
        title = "{Discovery of a Time Lag between the Soft X-Ray and Radio Emission of the Tidal Disruption Flare ASASSN-14li: Evidence for Linear Disk-Jet Coupling}",
      journal = {\apj},
         year = 2018,
        month = mar,
       volume = {856},
       number = {1},
          eid = {1},
        pages = {1},
          doi = {10.3847/1538-4357/aab361},
archivePrefix = {arXiv},
       eprint = {1709.02882},
 primaryClass = {astro-ph.HE},
       adsurl = {https://ui.adsabs.harvard.edu/abs/2018ApJ...856....1P}
}

@ARTICLE{Jin2025,
       author = {{Jin}, C. -C. and {Li}, D. -Y. and {Jiang}, N. and {Dai}, L. -X. and {Cheng}, H. -Q. and {Zhu}, J. -Z. and {Yang}, C. -W. and {Rau}, A. and {Baldini}, P. and {Wang}, T. -G. and {Zhou}, H. -Y. and {Yuan}, W. and {Zhang}, C. and {Shu}, X. -W. and {Shen}, R. -F. and {Wang}, Y. -L. and {Wen}, S. -X. and {Wu}, Q. -Y. and {Wang}, Y. -B. and {Thomsen}, L.~L. and {Zhang}, Z. -J. and {Zhang}, W. -J. and {Coleiro}, A. and {Eyles-Ferris}, R. and {Fang}, X. and {Ho}, L.~C. and {Hu}, J. -W. and {Jin}, J. -J. and {Li}, W. -X. and {Liu}, B. -F. and {Liu}, F. -K. and {Liu}, M. -J. and {Liu}, Z. and {Lu}, Y. -J. and {Merloni}, A. and {Qiao}, E. -L. and {Saxton}, R. and {Soria}, R. and {Wang}, S. and {Xue}, Y. -Q. and {Yang}, H. -N. and {Zhang}, B. and {Zhang}, W. -D. and {Cai}, Z. -M. and {Chen}, F. -S. and {Chen}, H. -L. and {Chen}, T. -X. and {Chen}, W. and {Chen}, Y. -H. and {Chen}, Y. -F. and {Chen}, Y. and {Cordier}, B. and {Cui}, C. -Z. and {Cui}, W. -W. and {Dai}, Y. -F. and {Ding}, H. -C. and {Fan}, D. -W. and {Fan}, Z. and {Feng}, H. and {Garcia}, J.~A. and {Guan}, J. and {Han}, D. -W. and {Hou}, D. -J. and {Hu}, H. -B. and {Huang}, M. -H. and {Huo}, J. and {Jia}, S. -M. and {Jia}, Z. -Q. and {Jiang}, B. -W. and {Jin}, G. and {Kong}, X. and {Kuulkers}, E. and {Lei}, W. -H. and {Li}, C. -K. and {Li}, J. -F. and {Li}, L. -H. and {Li}, M. -S. and {Li}, W. and {Li}, Z. -D. and {Lian}, T. -Y. and {Ling}, Z. -X. and {Liu}, C. -Z. and {Liu}, H. -Y and {Liu}, H. -Q. and {Liu}, J. -F. and {Liu}, Y. and {Lu}, F. -J. and {Luo}, L. -D. and {Ma}, J. and {Mao}, X. and {Mu}, H. -Y. and {Nandra}, K. and {O'Brien}, P. and {Pan}, H. -W. and {Pan}, X. and {Qin}, G. -J. and {Rea}, N. and {Sanders}, J. and {Song}, L. -M. and {Sun}, H. and {Sun}, S. -L. and {Sun}, X. -J. and {Tan}, Y. -Y. and {Tang}, Q. -J. and {Tao}, Y. -H. and {Wang}, B. -C. and {Wang}, J. and {Wang}, J. -F. and {Wang}, L. and {Wang}, W. -X. and {Wang}, Y. -S. and {Wang}, Z. -X. and {Wu}, Q. -W. and {Wu}, X. -F. and {Xu}, H. -T. and {Xu}, J. -J. and {Xu}, X. -P. and {Xu}, Y. -F. and {Xu}, Z. and {Xue}, C. -B. and {Xue}, S. -J. and {Xue}, Y. -L. and {Yan}, A. -L. and {Yang}, X. -T. and {Yang}, Y. -J. and {Zhang}, J. and {Zhang}, M. and {Zhang}, S. -N. and {Zhang}, Y. -H. and {Zhang}, Z. and {Zhang}, Z. and {Zhang}, Z. -L. and {Zhao}, D. -H. and {Zhao}, H. -S. and {Zhao}, X. -F. and {Zhao}, Z. -J. and {Zheng}, J. and {Zhu}, Q. -F. and {Zhu}, Y. -X. and {Zhu}, Z. -C. and {Zou}, H.},
        title = "{An Intermediate-mass Black Hole Lurking in A Galactic Halo Caught Alive during Outburst}",
      journal = {arXiv e-prints},
         year = 2025,
        month = jan,
          eid = {arXiv:2501.09580},
        pages = {arXiv:2501.09580},
          doi = {10.48550/arXiv.2501.09580},
archivePrefix = {arXiv},
       eprint = {2501.09580},
 primaryClass = {astro-ph.HE},
       adsurl = {https://ui.adsabs.harvard.edu/abs/2025arXiv250109580J}
}

@ARTICLE{alexander2020SSRv,
       author = {{Alexander}, Kate D. and {van Velzen}, Sjoert and {Horesh}, Assaf and {Zauderer}, B. Ashley},
        title = "{Radio Properties of Tidal Disruption Events}",
      journal = {\ssr},
         year = 2020,
        month = jun,
       volume = {216},
       number = {5},
          eid = {81},
        pages = {81},
          doi = {10.1007/s11214-020-00702-w},
archivePrefix = {arXiv},
       eprint = {2006.01159},
 primaryClass = {astro-ph.HE},
       adsurl = {https://ui.adsabs.harvard.edu/abs/2020SSRv..216...81A}
}

@incollection{TaoAn03.2026.SKA, author = {Tao An and author2 and author3 and author4 and author5},title = {},year = {2026},publisher = {},note = {arXiv search: Report number AASKAII/TaoAn03},booktitle = {Advancing Astrophysics with the SKA -- II (AASKAII)}}

@incollection{DAmato01.2026.SKA, author = {Quirino D'Amato and author2 and author3 and author4 and author5},title = {},year = {2026},publisher = {},note = {arXiv search: Report number AASKAII/DAmato01},booktitle = {Advancing Astrophysics with the SKA -- II (AASKAII)}}

@ARTICLE{Sokolovsky2011,
       author = {{Sokolovsky}, K.~V. and {Kovalev}, Y.~Y. and {Pushkarev}, A.~B. and {Lobanov}, A.~P.},
        title = "{A VLBA survey of the core shift effect in AGN jets. I. Evidence of dominating synchrotron opacity}",
      journal = {\aap},
         year = 2011,
        month = aug,
       volume = {532},
          eid = {A38},
        pages = {A38},
          doi = {10.1051/0004-6361/201016072},
archivePrefix = {arXiv},
       eprint = {1103.6032},
 primaryClass = {astro-ph.CO},
       adsurl = {https://ui.adsabs.harvard.edu/abs/2011A&A...532A..38S}
}

@ARTICLE{Yao2025,
       author = {{Yao}, Yuhan and {Chornock}, Ryan and {Ward}, Charlotte and {Hammerstein}, Erica and {Sfaradi}, Itai and {Margutti}, Raffaella and {Kelley}, Luke Zoltan and {Lu}, Wenbin and {Liu}, Chang and {Wise}, Jacob and {Sollerman}, Jesper and {Alexander}, Kate D. and {Bellm}, Eric C. and {Drake}, Andrew J. and {Fremling}, Christoffer and {Gilfanov}, Marat and {Graham}, Matthew J. and {Groom}, Steven L. and {Hinds}, K.~R. and {Kulkarni}, S.~R. and {Miller}, Adam A. and {Miller-Jones}, James C.~A. and {Nicholl}, Matt and {Perley}, Daniel A. and {Purdum}, Josiah and {Ravi}, Vikram and {Rich}, R. Michael and {Rehemtulla}, Nabeel and {Riddle}, Reed and {Smith}, Roger and {Stein}, Robert and {Sunyaev}, Rashid and {van Velzen}, Sjoert and {Wold}, Avery},
        title = "{A Massive Black Hole 0.8 kpc from the Host Nucleus Revealed by the Offset Tidal Disruption Event AT2024tvd}",
      journal = {\apjl},
         year = 2025,
        month = jun,
       volume = {985},
       number = {2},
          eid = {L48},
        pages = {L48},
          doi = {10.3847/2041-8213/add7de},
archivePrefix = {arXiv},
       eprint = {2502.17661},
 primaryClass = {astro-ph.GA},
       adsurl = {https://ui.adsabs.harvard.edu/abs/2025ApJ...985L..48Y}
}

@ARTICLE{Lu2018,
       author = {{Lu}, Wenbin and {Kumar}, Pawan},
        title = "{On the Missing Energy Puzzle of Tidal Disruption Events}",
      journal = {\apj},
         year = 2018,
        month = oct,
       volume = {865},
       number = {2},
          eid = {128},
        pages = {128},
          doi = {10.3847/1538-4357/aad54a},
archivePrefix = {arXiv},
       eprint = {1802.02151},
 primaryClass = {astro-ph.HE},
       adsurl = {https://ui.adsabs.harvard.edu/abs/2018ApJ...865..128L}
}

@ARTICLE{Beniamini2023,
       author = {{Beniamini}, Paz and {Piran}, Tsvi and {Matsumoto}, Tatsuya},
        title = "{Swift J1644+57 as an off-axis Jet}",
      journal = {\mnras},
         year = 2023,
        month = sep,
       volume = {524},
       number = {1},
        pages = {1386-1395},
          doi = {10.1093/mnras/stad1950},
archivePrefix = {arXiv},
       eprint = {2305.06370},
 primaryClass = {astro-ph.HE},
       adsurl = {https://ui.adsabs.harvard.edu/abs/2023MNRAS.524.1386B}
}

@ARTICLE{duran2013,
       author = {{Barniol Duran}, Rodolfo and {Nakar}, Ehud and {Piran}, Tsvi},
        title = "{Radius Constraints and Minimal Equipartition Energy of Relativistically Moving Synchrotron Sources}",
      journal = {\apj},
         year = 2013,
        month = jul,
       volume = {772},
       number = {1},
          eid = {78},
        pages = {78},
          doi = {10.1088/0004-637X/772/1/78},
archivePrefix = {arXiv},
       eprint = {1301.6759},
 primaryClass = {astro-ph.HE},
       adsurl = {https://ui.adsabs.harvard.edu/abs/2013ApJ...772...78B}
}

@ARTICLE{Trakhtenbrot2019b,
       author = {{Trakhtenbrot}, Benny and {Arcavi}, Iair and {MacLeod}, Chelsea L. and {Ricci}, Claudio and {Kara}, Erin and {Graham}, Melissa L. and {Stern}, Daniel and {Harrison}, Fiona A. and {Burke}, Jamison and {Hiramatsu}, Daichi and {Hosseinzadeh}, Griffin and {Howell}, D. Andrew and {Smartt}, Stephen J. and {Rest}, Armin and {Prieto}, Jose L. and {Shappee}, Benjamin J. and {Holoien}, Thomas W.-S. and {Bersier}, David and {Filippenko}, Alexei V. and {Brink}, Thomas G. and {Zheng}, WeiKang and {Li}, Ruancun and {Remillard}, Ronald A. and {Loewenstein}, Michael},
        title = "{1ES 1927+654: An AGN Caught Changing Look on a Timescale of Months}",
      journal = {\apj},
         year = 2019,
        month = sep,
       volume = {883},
       number = {1},
          eid = {94},
        pages = {94},
          doi = {10.3847/1538-4357/ab39e4},
archivePrefix = {arXiv},
       eprint = {1903.11084},
 primaryClass = {astro-ph.GA},
       adsurl = {https://ui.adsabs.harvard.edu/abs/2019ApJ...883...94T}
}

@ARTICLE{Ricci2020,
       author = {{Ricci}, C. and {Kara}, E. and {Loewenstein}, M. and {Trakhtenbrot}, B. and {Arcavi}, I. and {Remillard}, R. and {Fabian}, A.~C. and {Gendreau}, K.~C. and {Arzoumanian}, Z. and {Li}, R. and {Ho}, L.~C. and {MacLeod}, C.~L. and {Cackett}, E. and {Altamirano}, D. and {Gandhi}, P. and {Kosec}, P. and {Pasham}, D. and {Steiner}, J. and {Chan}, C.-H.},
        title = "{The Destruction and Recreation of the X-Ray Corona in a Changing-look Active Galactic Nucleus}",
      journal = {\apjl},
         year = 2020,
        month = jul,
       volume = {898},
       number = {1},
          eid = {L1},
        pages = {L1},
          doi = {10.3847/2041-8213/ab91a1},
archivePrefix = {arXiv},
       eprint = {2007.07275},
 primaryClass = {astro-ph.HE},
       adsurl = {https://ui.adsabs.harvard.edu/abs/2020ApJ...898L...1R}
}

@ARTICLE{Wu2022,
       author = {{Wu}, Han-Ji and {Mou}, Guobin and {Wang}, Kai and {Wang}, Wei and {Li}, Zhuo},
        title = "{Could TDE outflows produce the PeV neutrino events?}",
      journal = {\mnras},
         year = 2022,
        month = aug,
       volume = {514},
       number = {3},
        pages = {4406-4412},
          doi = {10.1093/mnras/stac1621},
archivePrefix = {arXiv},
       eprint = {2112.01748},
 primaryClass = {astro-ph.HE},
       adsurl = {https://ui.adsabs.harvard.edu/abs/2022MNRAS.514.4406W}
}

@ARTICLE{vanVelzen2024,
       author = {{van Velzen}, Sjoert and {Stein}, Robert and {Gilfanov}, Marat and {Kowalski}, Marek and {Hayasaki}, Kimitake and {Reusch}, Simeon and {Yao}, Yuhan and {Garrappa}, Simone and {Franckowiak}, Anna and {Gezari}, Suvi and {Nordin}, Jakob and {Fremling}, Christoffer and {Sharma}, Yashvi and {Yan}, Lin and {Kool}, Erik C. and {Stern}, Daniel and {Veres}, Patrik M. and {Sollerman}, Jesper and {Medvedev}, Pavel and {Sunyaev}, Rashid and {Bellm}, Eric C. and {Dekany}, Richard G. and {Duev}, Dimitri A. and {Graham}, Matthew J. and {Kasliwal}, Mansi M. and {Kulkarni}, Shrinivas R. and {Laher}, Russ R. and {Riddle}, Reed L. and {Rusholme}, Ben},
        title = "{Establishing accretion flares from supermassive black holes as a source of high-energy neutrinos}",
      journal = {\mnras},
         year = 2024,
        month = apr,
       volume = {529},
       number = {3},
        pages = {2559-2576},
          doi = {10.1093/mnras/stae610},
archivePrefix = {arXiv},
       eprint = {2111.09391},
 primaryClass = {astro-ph.HE},
       adsurl = {https://ui.adsabs.harvard.edu/abs/2024MNRAS.529.2559V}
}

@ARTICLE{Liu2014,
       author = {{Liu}, F.~K. and {Li}, Shuo and {Komossa}, S.},
        title = "{A Milliparsec Supermassive Black Hole Binary Candidate in the Galaxy SDSS J120136.02+300305.5}",
      journal = {\apj},
         year = 2014,
        month = may,
       volume = {786},
       number = {2},
          eid = {103},
        pages = {103},
          doi = {10.1088/0004-637X/786/2/103},
archivePrefix = {arXiv},
       eprint = {1404.4933},
 primaryClass = {astro-ph.HE},
       adsurl = {https://ui.adsabs.harvard.edu/abs/2014ApJ...786..103L}
}

@ARTICLE{Shu2020,
       author = {{Shu}, Xinwen and {Zhang}, Wenjie and {Li}, Shuo and {Jiang}, Ning and {Dou}, Liming and {Yan}, Zhen and {Xie}, Fu-Guo and {Shen}, Rongfeng and {Sun}, Luming and {Liu}, Fukun and {Wang}, Tinggui},
        title = "{X-ray flares from the stellar tidal disruption by a candidate supermassive black hole binary}",
      journal = {Nature Communications},
         year = 2020,
        month = nov,
       volume = {11},
          eid = {5876},
        pages = {5876},
          doi = {10.1038/s41467-020-19675-z},
archivePrefix = {arXiv},
       eprint = {2012.11818},
 primaryClass = {astro-ph.HE},
       adsurl = {https://ui.adsabs.harvard.edu/abs/2020NatCo..11.5876S}
}

@ARTICLE{mou2025,
       author = {{Mou}, Guobin},
        title = "{Numerical Studies on the Radio Afterglows in TDE (I): Forward Shock}",
      journal = {arXiv e-prints},
         year = 2025,
        month = oct,
          eid = {arXiv:2510.14715},
        pages = {arXiv:2510.14715},
archivePrefix = {arXiv},
       eprint = {2510.14715},
 primaryClass = {astro-ph.HE},
       adsurl = {https://ui.adsabs.harvard.edu/abs/2025arXiv251014715M}
}

@ARTICLE{moushu2025,
       author = {{Mou}, Guobin and {Shu}, Xinwen},
        title = "{Numerical Studies on the Radio Afterglows in TDE: Bow Shock}",
      journal = {arXiv e-prints},
         year = 2025,
        month = oct,
          eid = {arXiv:2510.25033},
        pages = {arXiv:2510.25033},
archivePrefix = {arXiv},
       eprint = {2510.25033},
 primaryClass = {astro-ph.HE},
       adsurl = {https://ui.adsabs.harvard.edu/abs/2025arXiv251025033M},
}

@ARTICLE{Stein2021,
       author = {{Stein}, Robert and {van Velzen}, Sjoert and {Kowalski}, Marek and {Franckowiak}, Anna and {Gezari}, Suvi and {Miller-Jones}, James C.~A. and {Frederick}, Sara and {Sfaradi}, Itai and {Bietenholz}, Michael F. and {Horesh}, Assaf and {Fender}, Rob and {Garrappa}, Simone and {Ahumada}, Tom{\'a}s and {Andreoni}, Igor and {Belicki}, Justin and {Bellm}, Eric C. and {B{\"o}ttcher}, Markus and {Brinnel}, Valery and {Burruss}, Rick and {Cenko}, S. Bradley and {Coughlin}, Michael W. and {Cunningham}, Virginia and {Drake}, Andrew and {Farrar}, Glennys R. and {Feeney}, Michael and {Foley}, Ryan J. and {Gal-Yam}, Avishay and {Golkhou}, V. Zach and {Goobar}, Ariel and {Graham}, Matthew J. and {Hammerstein}, Erica and {Helou}, George and {Hung}, Tiara and {Kasliwal}, Mansi M. and {Kilpatrick}, Charles D. and {Kong}, Albert K.~H. and {Kupfer}, Thomas and {Laher}, Russ R. and {Mahabal}, Ashish A. and {Masci}, Frank J. and {Necker}, Jannis and {Nordin}, Jakob and {Perley}, Daniel A. and {Rigault}, Mickael and {Reusch}, Simeon and {Rodriguez}, Hector and {Rojas-Bravo}, C{\'e}sar and {Rusholme}, Ben and {Shupe}, David L. and {Singer}, Leo P. and {Sollerman}, Jesper and {Soumagnac}, Maayane T. and {Stern}, Daniel and {Taggart}, Kirsty and {van Santen}, Jakob and {Ward}, Charlotte and {Woudt}, Patrick and {Yao}, Yuhan},
        title = "{A tidal disruption event coincident with a high-energy neutrino}",
      journal = {Nature Astronomy},
         year = 2021,
        month = feb,
       volume = {5},
        pages = {510-518},
          doi = {10.1038/s41550-020-01295-8},
archivePrefix = {arXiv},
       eprint = {2005.05340},
 primaryClass = {astro-ph.HE},
       adsurl = {https://ui.adsabs.harvard.edu/abs/2021NatAs...5..510S}
}

@ARTICLE{Prieto2024,
       author = {{Prieto}, Almudena and {Gladis Magris}, C. and {Bruzual}, Gustavo and {Fern{\'a}ndez-Ontiveros}, Juan A. and {Burkert}, Andreas},
        title = "{The PARSEC view of star formation in galaxy centres: from protoclusters to star clusters in an early-type spiral}",
      journal = {\mnras},
         year = 2024,
        month = sep,
       volume = {533},
       number = {1},
        pages = {433-454},
          doi = {10.1093/mnras/stae1822},
archivePrefix = {arXiv},
       eprint = {2407.20860},
 primaryClass = {astro-ph.GA},
       adsurl = {https://ui.adsabs.harvard.edu/abs/2024MNRAS.533..433P}
}

@ARTICLE{KomossaMerritt2008,
       author = {{Komossa}, Stefanie and {Merritt}, David},
        title = "{Tidal Disruption Flares from Recoiling Supermassive Black Holes}",
      journal = {\apjl},
         year = 2008,
        month = aug,
       volume = {683},
       number = {1},
        pages = {L21},
          doi = {10.1086/591420},
archivePrefix = {arXiv},
       eprint = {0807.0223},
 primaryClass = {astro-ph},
       adsurl = {https://ui.adsabs.harvard.edu/abs/2008ApJ...683L..21K}
}

@ARTICLE{LiuLi2009,
       author = {{Liu}, F.~K. and {Li}, S. and {Chen}, Xian},
        title = "{Interruption of Tidal-Disruption Flares by Supermassive Black Hole Binaries}",
      journal = {\apjl},
         year = 2009,
        month = nov,
       volume = {706},
       number = {1},
        pages = {L133-L137},
          doi = {10.1088/0004-637X/706/1/L133},
archivePrefix = {arXiv},
       eprint = {0910.4152},
 primaryClass = {astro-ph.HE},
       adsurl = {https://ui.adsabs.harvard.edu/abs/2009ApJ...706L.133L}
}

@ARTICLE{Lin2020,
       author = {{Lin}, Dacheng and {Strader}, Jay and {Romanowsky}, Aaron J. and {Irwin}, Jimmy A. and {Godet}, Olivier and {Barret}, Didier and {Webb}, Natalie A. and {Homan}, Jeroen and {Remillard}, Ronald A.},
        title = "{Multiwavelength Follow-up of the Hyperluminous Intermediate-mass Black Hole Candidate 3XMM J215022.4-055108}",
      journal = {\apjl},
         year = 2020,
        month = apr,
       volume = {892},
       number = {2},
          eid = {L25},
        pages = {L25},
          doi = {10.3847/2041-8213/ab745b},
archivePrefix = {arXiv},
       eprint = {2002.04618},
 primaryClass = {astro-ph.HE},
       adsurl = {https://ui.adsabs.harvard.edu/abs/2020ApJ...892L..25L}
}

@ARTICLE{Wei2016,
       author = {{Wei}, J. and {Cordier}, B. and {Antier}, S. and {Antilogus}, P. and {Atteia}, J. -L. and {Bajat}, A. and {Basa}, S. and {Beckmann}, V. and {Bernardini}, M.~G. and {Boissier}, S. and {Bouchet}, L. and {Burwitz}, V. and {Claret}, A. and {Dai}, Z. -G. and {Daigne}, F. and {Deng}, J. and {Dornic}, D. and {Feng}, H. and {Foglizzo}, T. and {Gao}, H. and {Gehrels}, N. and {Godet}, O. and {Goldwurm}, A. and {Gonzalez}, F. and {Gosset}, L. and {G{\"o}tz}, D. and {Gouiffes}, C. and {Grise}, F. and {Gros}, A. and {Guilet}, J. and {Han}, X. and {Huang}, M. and {Huang}, Y. -F. and {Jouret}, M. and {Klotz}, A. and {La Marle}, O. and {Lachaud}, C. and {Le Floch}, E. and {Lee}, W. and {Leroy}, N. and {Li}, L. -X. and {Li}, S.~C. and {Li}, Z. and {Liang}, E. -W. and {Lyu}, H. and {Mercier}, K. and {Migliori}, G. and {Mochkovitch}, R. and {O'Brien}, P. and {Osborne}, J. and {Paul}, J. and {Perinati}, E. and {Petitjean}, P. and {Piron}, F. and {Qiu}, Y. and {Rau}, A. and {Rodriguez}, J. and {Schanne}, S. and {Tanvir}, N. and {Vangioni}, E. and {Vergani}, S. and {Wang}, F. -Y. and {Wang}, J. and {Wang}, X. -G. and {Wang}, X. -Y. and {Watson}, A. and {Webb}, N. and {Wei}, J.~J. and {Willingale}, R. and {Wu}, C. and {Wu}, X. -F. and {Xin}, L. -P. and {Xu}, D. and {Yu}, S. and {Yu}, W. -F. and {Yu}, Y. -W. and {Zhang}, B. and {Zhang}, S. -N. and {Zhang}, Y. and {Zhou}, X.~L.},
        title = "{The Deep and Transient Universe in the SVOM Era: New Challenges and Opportunities - Scientific prospects of the SVOM mission}",
      journal = {arXiv e-prints},
         year = 2016,
        month = oct,
          eid = {arXiv:1610.06892},
        pages = {arXiv:1610.06892},
          doi = {10.48550/arXiv.1610.06892},
archivePrefix = {arXiv},
       eprint = {1610.06892},
 primaryClass = {astro-ph.IM},
       adsurl = {https://ui.adsabs.harvard.edu/abs/2016arXiv161006892W}
}

@ARTICLE{Yuan2025,
       author = {{Yuan}, Weimin and {Dai}, Lixin and {Feng}, Hua and {Jin}, Chichuan and {Jonker}, Peter and {Kuulkers}, Erik and {Liu}, Yuan and {Nandra}, Kirpal and {O'Brien}, Paul and {Piro}, Luigi and {Rau}, Arne and {Rea}, Nanda and {Sanders}, Jeremy and {Tao}, Lian and {Wang}, Junfeng and {Wu}, Xuefeng and {Zhang}, Bing and {Zhang}, Shuangnan and {Ai}, Shunke and {Buchner}, Johannes and {Bulbul}, Esra and {Chen}, Hechao and {Chen}, Minghua and {Chen}, Yong and {Chen}, Yu-Peng and {Coleiro}, Alexis and {Zelati}, Francesco Coti and {Dai}, Zigao and {Fan}, Xilong and {Fan}, Zhou and {Friedrich}, Susanne and {Gao}, He and {Ge}, Chong and {Ge}, Mingyu and {Geng}, Jinjun and {Ghirlanda}, Giancarlo and {Gianfagna}, Giulia and {Gou}, Lijun and {Guillot}, S{\'e}bastien and {Hou}, Xian and {Hu}, Jingwei and {Huang}, Yongfeng and {Ji}, Long and {Jia}, Shumei and {Komossa}, S. and {Kong}, Albert K.~H. and {Lan}, Lin and {Li}, An and {Li}, Ang and {Li}, Chengkui and {Li}, Dongyue and {Li}, Jian and {Li}, Zhaosheng and {Ling}, Zhixing and {Liu}, Ang and {Liu}, Jinzhong and {Liu}, Liangduan and {Liu}, Zhu and {Luo}, Jiawei and {Ma}, Ruican and {Maggi}, Pierre and {Maitra}, Chandreyee and {Marino}, Alessio and {Ng}, Stephen Chi-Yung and {Pan}, Haiwu and {Rukdee}, Surangkhana and {Soria}, Roberto and {Sun}, Hui and {Tam}, Pak-Hin Thomas and {Thakur}, Aishwarya Linesh and {Tian}, Hui and {Troja}, Eleonora and {Wang}, Wei and {Wang}, Xiangyu and {Wang}, Yanan and {Wei}, Junjie and {Wen}, Sixiang and {Wu}, Jianfeng and {Wu}, Ting and {Xiao}, Di and {Xu}, Dong and {Xu}, Renxin and {Xu}, Yanjun and {Xu}, Yu and {Yang}, Haonan and {You}, Bei and {Yu}, Heng and {Yu}, Yunwei and {Zhang}, Binbin and {Zhang}, Chen and {Zhang}, Guobao and {Zhang}, Liang and {Zhang}, Wenda and {Zhang}, Yu and {Zhou}, Ping and {Zou}, Zecheng},
        title = "{Science objectives of the Einstein Probe mission}",
      journal = {Science China Physics, Mechanics, and Astronomy},
         year = 2025,
        month = mar,
       volume = {68},
       number = {3},
          eid = {239501},
        pages = {239501},
          doi = {10.1007/s11433-024-2600-3},
archivePrefix = {arXiv},
       eprint = {2501.07362},
 primaryClass = {astro-ph.HE},
       adsurl = {https://ui.adsabs.harvard.edu/abs/2025SCPMA..6839501Y}
}

@ARTICLE{Yang2016,
       author = {{Yang}, J. and {Paragi}, Z. and {van der Horst}, A.~J. and {Gurvits}, L.~I. and {Campbell}, R.~M. and {Giannios}, D. and {An}, T. and {Komossa}, S.},
        title = "{No apparent superluminal motion in the first-known jetted tidal disruption event Swift J1644+5734}",
      journal = {\mnras},
         year = 2016,
        month = oct,
       volume = {462},
       number = {1},
        pages = {L66-L70},
          doi = {10.1093/mnrasl/slw107},
archivePrefix = {arXiv},
       eprint = {1605.06461},
 primaryClass = {astro-ph.HE},
       adsurl = {https://ui.adsabs.harvard.edu/abs/2016MNRAS.462L..66Y}
}

\end{document}